\documentclass[twocolumn,preprintnumbers,amsmath,amssymb,prx,nolongbibliography]{revtex4-2}

\usepackage{physics}
\usepackage{graphicx}
\usepackage{color}
\usepackage{hyperref}
\usepackage{amsfonts}

\usepackage{circuitikz}
\usetikzlibrary{shapes.geometric, arrows.meta}

\hypersetup{
colorlinks = true,
urlcolor = blue,
linkcolor = blue,
citecolor = blue
}

\begin{document}

\title{ Passive memories beyond the Ising model: from SFQ to GKP}

\author{Simon Lieu}
\email{slieu@amazon.com}
\affiliation{AWS Center for Quantum Computing, Pasadena, California 91125, USA}
\date{\today}

\begin{abstract}

Self-correcting quantum memories are spin Hamiltonians that aim to generalize the classical memory of the magnetic hard disk (i.e.~ferromagnetic Ising model) to protect against both bit flips and phase flips.
Beyond the magnetic hard disk, there are other passive  classical memories that are used in modern computers, notably:  the flash memory in laptops (which traps charge on a conducting island), and the flux storage loop in single-flux quantum (SFQ)  logic (which traps flux in a superconducting loop).  
In this paper we review how storage and writing work in these two classical
memories, drawing analogies to the charge  qubit and the flux qubit.
Protection in these systems derives not from
scaling a lattice to the thermodynamic limit, but from a  large, tunable energy barrier that separates codewords
during memory, yet can be suppressed during a write.
Towards protecting a qubit: We introduce a 2D ``floating XY model'' that  combines the exponential (phase-flip) protection of the floating charge island with the polynomial (bit-flip) protection of the ferromagnetic XY model. We also study a  self-dual superconducting circuit that  hosts GKP codewords as symmetry-broken ground states of a 
 $\mathbb{Z}_2 \times \mathbb{Z}_2  $-symmetric circuit, and discuss its protection and limitations.
\end{abstract}

\maketitle

\section{Introduction}

Building a computer is challenging because it requires  both robust storage and control of information (bits/qubits). To store information, we need to encode it in a physical system and make sure that environmental perturbations do not alter it; to control it, we need our signals to be able to couple to the system strongly enough to alter its state.  Satisfying these conflicting requirements is particularly challenging in quantum computers.  In this work, we take a step back and revisit the classical problem: How  do  actual classical memories store and write bits so robustly? While indeed the classical problem is much easier than the quantum one (because one does not need to worry about phase flips), it is still challenging and we might learn lessons that are applicable to the quantum case.

Historically, the field of self-correcting quantum memories  has focused on generalizing the protection found in the  magnetic hard disk (modeled via a ferromagnetic  Ising model) to its quantum counterpart, i.e.~finding a quantum hard drive \cite{eric2002, brown2016, yoshida2011,  terhal2015, bombin2013, haah2011, bravyi2013, roberts2020b,  Alicki2010,  pastawski2011,  brell2014, bacon2006, lake2024, gu2025layer, williamson2025partial,roberts2026, 
placke2024, hong2025, sang2024, tibor2024, kumar2026, guo2026, lin2024, hsin2025, stahl2021,liu2023dissipative}.  The Ising model has been the gold standard 
of a classical memory for several reasons: It is physically realizable, stable in thermal equilibrium,  and has a simple Hamiltonian description that can be generalized to the quantum case.  The approach has been: Start by identifying the rules that the Ising model obeys \cite{brell16} (to benchmark our intuition for what is physically reasonable), then try to find a  quantum spin Hamiltonian that  can analogously protect both bit and phase flips, e.g.~via extensive energy barriers between codewords. This approach has led to a number of interesting models for passive quantum memories, including the 4D toric code \cite{eric2002, Alicki2010}, fractons \cite{haah2011},  and non-translationally-invariant models \cite{brell16, balasubramanian2026passive}. However, it remains challenging to find natural systems (e.g. materials) or engineered systems (e.g.  lattices of superconducting qubits) that obey such interactions.

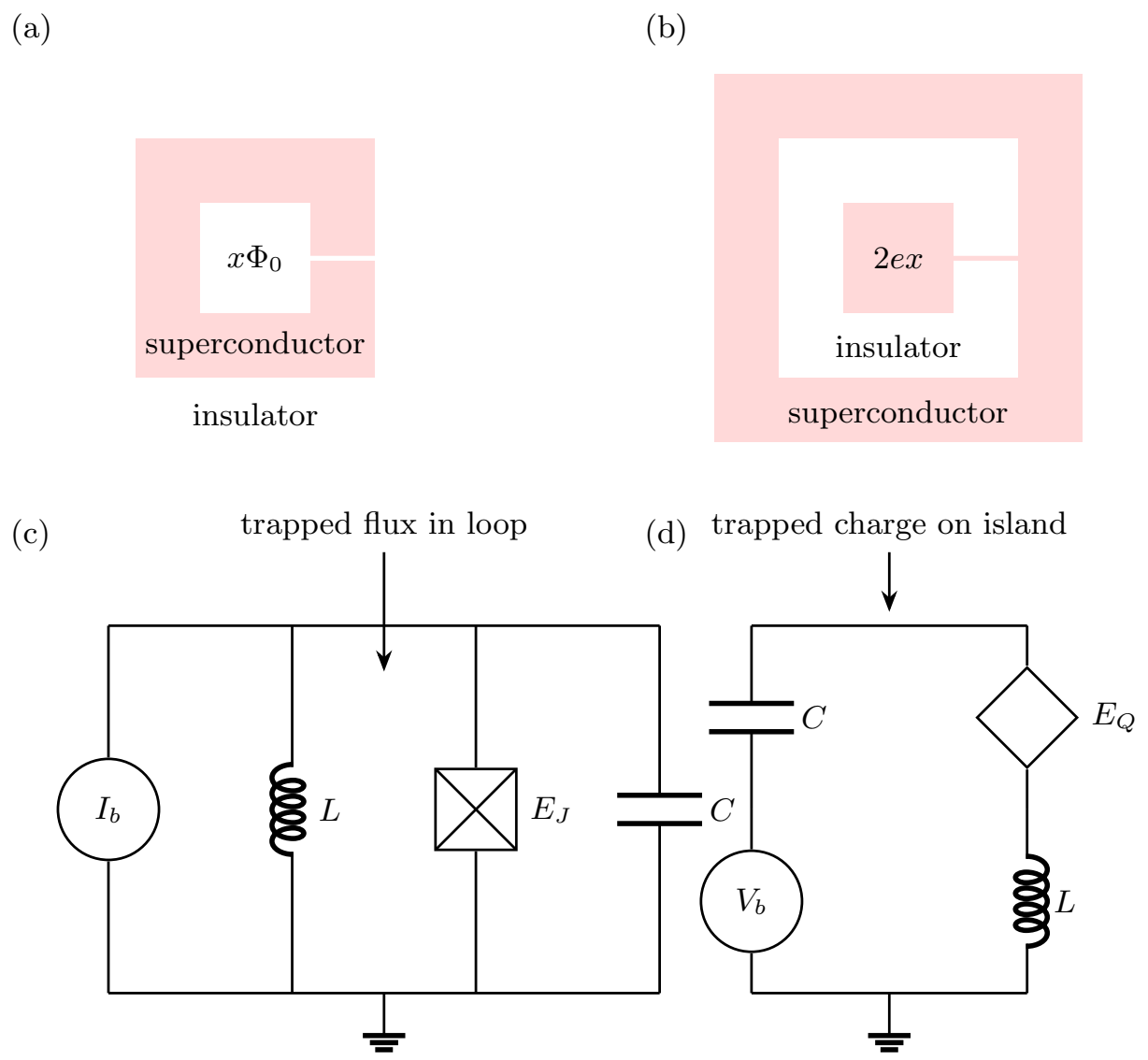
\begin{figure}[!t]
    \centering
\def\figtextscale{1.1} 
\scalebox{0.65}{%

\begin{circuitikz}[thick, every node/.style={scale=\figtextscale}]
\begin{scope}[shift={(1.6,8)}]

\fill[white] (-2,-2) rectangle (2,2);
\fill[red!15] (-1.3,-1.3) rectangle (1.3,1.3);
\fill[white] (-0.6,-0.6) rectangle (0.6,0.6);

\draw[white, line width=1.5pt] (0,0) -- (2,0);

\node at (0,-0.95) {\small superconductor};

\node at (0,0) {$x\Phi_0$};

\node at (0,-1.7) {\small insulator};

\node[anchor=north west] at (-2.8,2.8) { (a)};

\end{scope}

\begin{scope}[shift={(8.6,8)}]

\fill[red!15] (-2,-2) rectangle (2,2);
\fill[white] (-1.3,-1.3) rectangle (1.3,1.3);

\fill[red!15] (-0.6,-0.6) rectangle (0.6,0.6);

\draw[red!15, line width=1.5pt] (0,0) -- (2,0);

\node at (0,-0.95) {\small insulator};

\node at (0,0) {$2ex$};

\node at (0,-1.7) {\small superconductor};

\node[anchor=north west] at (-2.9,2.8) {(b)};

\end{scope}

\begin{scope}[shift={(0,0)}]

\node[draw, circle, minimum size=1cm, inner sep=0pt] (I) at (0,2) {$I_b$};
\draw (0,0) -- (I.south);
\draw (I.north) -- (0,4);

\draw (2,0) to[L] (2,4);
\node[right, xshift=4pt] at (2,2) {$L$};

\node[draw, rectangle, minimum size=0.8cm, inner sep=0pt,
  path picture={
    \draw (path picture bounding box.south west) -- (path picture bounding box.north east);
    \draw (path picture bounding box.north west) -- (path picture bounding box.south east);
  }] (JJ) at (4,2) {};
\draw (4,0) -- (JJ.south);
\draw (JJ.north) -- (4,4);

\draw (6,0) to[C] (6,4);
\node[right, xshift=8pt] at (6.1,2) {$C$};

\draw (0,4) -- (6,4);
\draw (0,0) -- (6,0);

\node[ground] at (3,0) {};

\node[right] at (JJ.east) {$E_J$};

\draw[-{Stealth}, thick] (3,4.8) node[above] {trapped flux in loop} -- (3,3.5);

\node[anchor=north west] at (-1.2,5.3) {(c)};

\end{scope}

\begin{scope}[shift={(7,0)}]

\draw (0,4) to[C] (0,2);
\node[right, xshift=8pt] at (0.1,3) {$C$};
\node[draw, circle, minimum size=1cm, inner sep=0pt] (V) at (0,1) {$V_b$};
\draw (0,2) -- (V.north);
\draw (V.south) -- (0,0);

\node[draw, diamond, minimum size=1cm, inner sep=0pt] (QPS) at (3,3) {};
\draw (3,4) -- (QPS.north);
\draw (QPS.south) -- (3,2);
\draw (3,2) to[L] (3,0);
\node[right, xshift=4pt] at (3,1) {$L$};

\draw (0,4) -- (3,4);
\draw (0,0) -- (3,0);

\node[ground] at (1.5,0) {};

\node[right] at (QPS.east) {$E_Q$};

\draw[-{Stealth}, thick] (1.5,4.8) node[above] {trapped charge on island} -- (1.5,4.15);

\node[anchor=north west] at (-1.3,5.3) {(d)};

\end{scope}

\end{circuitikz}}

    \caption{  (a) Classical memory in SFQ storage:
    A flux quantum can be trapped in the middle of a superconducting loop (with a Josephson junction), storing classical information in $x =0,1$. (b)
    Classical memory in the (``noncompact'') Cooper pair box [electromagnetic dual of (a)]: A superconducting island is surrounded by an insulator [with a thin superconducting wire to ground (quantum phase slip junction)]. A Cooper pair can be  trapped on the island, storing classical information in $x =0,1$.   (c) Lumped element circuit diagram for (a): to write to the flux memory, one applies a  current pulse ($I_b$) to inject/remove a flux quantum from the loop.
     (d) Lumped element circuit diagram for (b): to write to the island, one applies a  voltage pulse ($V_b$) to inject/remove a Cooper pair from the island.}
    \label{fig:classical_memory_cartoon}
\end{figure}

Beyond the magnetic hard disk,  there are other  passive classical  memories  that cannot be described via the model of ``a lattice of noisy two-level systems'', and it is worth playing the same game with those: Start with uncovering the rules of the classical protection then try to generalize the models to host   quantum protection. In this paper we  look at how  storing and  writing to  a  passive classical memory works in practice, focusing on two separate platforms: the flash memory in the solid-state drive (used in modern laptops) \cite{sze2007physics, pavan1997flash, aritome_2015} and the flux storage loop [used in single-flux quantum (SFQ) logic \cite{likharev1991rsfq, mukhanov2011, Takeuchi_2013} - a leading candidate for the classical control hardware of superconducting qubits \cite{mcdermott_2019, howe_2022, ferguson_2020, ng_2026, ng_2026b}].  Both of these  memories are (like the Ising model) physically realizable and stable in thermal equilibrium.  What are the mechanisms that protect the bit in these systems? And can we find an analogous quantum memory? We wish to explain the protection in a language that is compatible with the modern field of superconducting qubits.  [See Fig.~\ref{fig:classical_memory_cartoon}.]

This work aims to  view the fields of self-correcting quantum memories \cite{eric2002, brown2016, yoshida2011,  terhal2015, bombin2013, haah2011, bravyi2013, roberts2020b,  Alicki2010,  pastawski2011,  brell2014, bacon2006, lake2024, gu2025layer, williamson2025partial,roberts2026, 
placke2024, hong2025, sang2024, tibor2024, kumar2026, guo2026, lin2024, hsin2025, stahl2021,liu2023dissipative} and protected superconducting qubits \cite{schuster2021, Kalashnikov2020,ferguson_2014, Smith2020, larsen2020, Rymarz2021, gyenis2021, Smith2022, leroux2023, Ciaccia2024, xanda_2024, esencan2026, Kumar2024, Leblanc2025, Smith2025, Hays2025, nguyen2025, Kolesnikow2026,siegele2026, Shagalov2026, Messelot2026, Roverch2026, Aramburu2026, lieu_2025, putterman2024} as two distinct approaches towards the same goal: How do we build a physical Hamiltonian whose degenerate qubit ground states are intrinsically protected against noise? And can  we start to combine ideas from these two fields to  benefit from each other by forming a lattice of superconducting qubits \cite{ioffe_2012, kitaev2006_mirror, epstein2022, gyenis_2022, dodge_2023b, lieu2023candidate, lieu_2025}?  
We discuss the ``rules'' of protection in superconducting qubits  and  how they differ from the typical ones imposed on self-correcting quantum memories of two-level systems.   The punchline is that SFQ and flash memories are protected not by lattice size but by an engineered separation of energy scales, e.g.~in SFQ, the bit-flip time scales as: $T_{bf}\sim \exp[c E_J/ (k_B T)]$ where $E_J$ is an explicit energy scale in the Hamiltonian that can be much larger than $k_B T$.
Towards protecting a full qubit,  we introduce a floating XY model  that combines the exponential (phase-flip) protection of the floating charge island with the polynomial (bit-flip) protection of the 2D ferromagnetic  XY model. We also study a self-dual superconducting circuit that  hosts Gottesman-Kitaev-Preskill (GKP) codewords as ground states. We show that the ground states obey an interesting symmetry breaking pattern of a $\mathbb{Z}_2 \times \mathbb{Z}_2$ symmetry whereby each codeword pointer state is symmetric with respect to one of the $ \mathbb{Z}_2$ symmetries but breaks the other. We argue that the model can protect against noise that is local in phase space but that realistic circuit implementations also host nonlocal noise that may limit  performance.

\section{Classical memory via trapped flux}

We begin by describing  a minimal version of the memory in single-flux quantum (SFQ) logic devices \cite{likharev1991rsfq, mukhanov2011, Takeuchi_2013}: the flux storage loop. [See Fig.~\ref{fig:classical_memory_cartoon}(a,c)]. SFQ is a family of superconducting circuits that are designed to do  classical memory and computation, typically by sending temporally sharp voltage pulses ($\sim$ps width) through Josephson transmission lines (a ladder of inductors and Josephson junctions).  Its main advantages compared to standard CMOS technologies are its high clock rate   and its low power consumption. A disadvantage is that it  requires long arrays of Josephson junctions which generally need to be fabricated with high yield.  SFQ devices have shown much promise  as an application specific integrated circuit for  the classical control hardware of superconducting qubits, i.e.~in performing a limited set of repeated tasks at high speed and low power dissipation \cite{mcdermott_2019, howe_2022, ferguson_2020, ng_2026, ng_2026b}.

Consider the superconducting circuit in Fig.~\ref{fig:classical_memory_cartoon}(c), i.e.~a current source, inductor, Josephson junction, and capacitor all in parallel (as well as a resistor also in parallel which is not shown).
Unlike  superconducting qubits, SFQ circuits generally operate in a regime where the inductive energies dominate over the capacitive ones: $E_J, E_L \gg E_C$ (where $E_C = e^2/(2C)$ is the charging energy and $E_L = (\Phi_0/2\pi)^2/L$ is the inductive energy), such that the superconducting phase degree of freedom $\phi$ behaves as
a classical variable: the quantum state remains a wavepacket sharply
peaked around a central value, with phase spread
$\Delta\phi \sim [2E_C/(E_J)]^{1/4} \ll 2\pi$, whose center
follows the classical trajectory, $\langle\phi\rangle(t) = \phi(t)$.
A typical set of parameters for the storage loop is: $E_J/h = 30 \text{ THz}, E_L/h = 10 \text{ THz}, E_C/h = 100 \text{ MHz}$, so junction and inductive energies that are about 3 orders of magnitude larger than in superconducting qubits.  In this parameter regime, the full Hamiltonian (or Lindbladian) description of the model is more complicated than necessary, and one generally works with Kirchhoff's laws for the dynamics of the $\phi$.  For the circuit above, Kirchhoff's law of conserved current leads to the following equation:
\begin{equation}\label{eq:kirchoff_current}
\left( \frac{\hbar}{2e} \right) I_b
= E_J \sin\phi
+ \left( \frac{\hbar^2}{4e^2 R} \right)\dot\phi
+ \left( \frac{\hbar^2}{8 E_C}\right) \ddot\phi
+ E_L (\phi - \pi),
\end{equation}
where we assume a constant DC current bias $I_{b,0} = (2e/\hbar) \pi E_L$  responsible for the $- E_L  \pi$ term, and  $I_b$ is any bias current above that, i.e.~$I_{b,\text{tot} } =I_{b,0}  + I_b $; $R$ is the resistance. The current through the current source must be equal to the sum of the current through each other element. We can gain intuition on this dynamics by defining the potential energy:
\begin{equation}
U(\phi) = -E_J \cos\phi + \frac{E_L}{2} (\phi - \pi)^2 - \left( \frac{\hbar}{2e} \right) I_b \phi.
\end{equation}
Kirchhoff's current law can then be expressed as:
\begin{equation}
-\frac{\partial U}{\partial \phi} = \left( \frac{\hbar^2}{8 E_C} \right) \ddot\phi
+ \left( \frac{\hbar^2}{4e^2 R} \right) \dot\phi,
\end{equation}
which is the classical equation of motion of a particle with mass $m= \hbar^2 / (8 E_C)$ in a potential $U$ with damping friction $\Gamma = \hbar^2/(4e^2 R)$.

\begin{figure}
    \centering
    \includegraphics[scale=0.4]{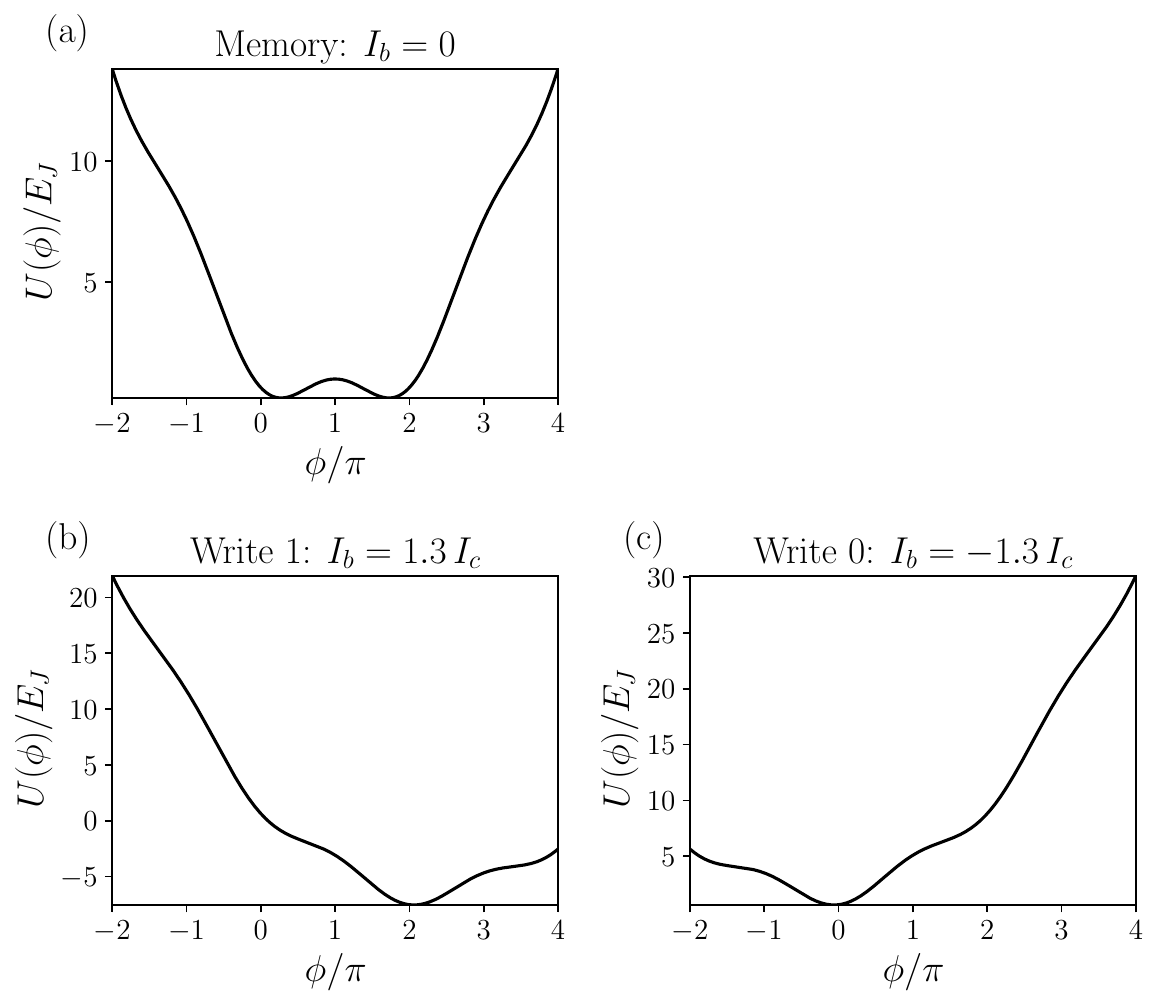}
    \caption{ The potential $U(\phi)$ during (a) memory, (b) writing to 1, (c) writing to 0. [$E_J/E_L = 3.$] The state is modeled as a classical ball that is trapped in one of the two wells in memory, but deterministically evolves to one of them during a write.  }
    \label{fig:sfq_potential}
\end{figure}

When $I_b=0$, the potential is that of the flux qubit, i.e.~a sum of a cosine and parabola depicted in Fig.~\ref{fig:sfq_potential}(a). The particle is trapped in a degenerate minimum of the potential, near the points $\phi = 0,2\pi$. These different minima physically correspond to different persistent current configurations, i.e.~counterclockwise current resulting in total flux $\Phi_0$ through the inductive loop, or clockwise current resulting in  zero total flux.
When $I_b=0$, these states are very stable: The barrier height between the two minima is approximately $E_J /h = 30 \text{ THz} \gg k_B T/h = 84 \text{ GHz}$ at $T= 4 \text{ K}$. The approximate bit-flip time of thermal hopping from one well to its neighbor obeys the Arrhenius law: $T_{bf} \sim \exp[ c E_J / (k_B T) ] \sim \exp[  357 c]$   \cite{lieu_2025}, which is practically infinite \footnote{The Arrhenius form holds above the quantum crossover temperature $k_B T^* \sim \hbar\omega_p/(2\pi)$, with $\omega_p = \sqrt{8E_JE_C}/\hbar$ the plasma frequency; below it the rate is set by macroscopic quantum tunneling, with a temperature-independent exponent $\sim\sqrt{E_J/E_C}$ that is similarly enormous here.}.

Given that the persistent current states in the flux storage loop are so stable to environmental noise, how does one change the state  to write to the memory?  
Suppose we  diabatically quench the circuit with a  pulse $I_b > I_c$. ($I_c = 2e E_J/\hbar$ is the critical current.)   Then the double-well potential gets deformed to a single-well potential centered near $\phi=2\pi$ for $I_b > I_c$, and near $\phi=0$ for $I_b <-I_c$. [See Fig.~\ref{fig:sfq_potential}(b,c).] Regardless of the starting point, the state   deterministically falls to one side causing a write to 0 or  1.  Importantly, there needs to be some resistance in order to damp the particle otherwise it would ring forever around the new minimum. The circuit parameters that typically satisfy these requirements are: $\beta_L =E_J/E_L\sim 3, \beta_R = 2\pi I_c R^2 C/ \Phi_0 \sim 0.5$.
The integral of the voltage satisfies: $\int V(t)dt = \Phi_0 (\phi_{end} - \phi_{start})/(2\pi) \approx \pm \Phi_0$ if the particle hopped wells, and $=0$ if it did not. Hopping wells is equivalent to injecting/removing one flux quantum  from the loop.

In practice, this current pulse is made by an SFQ pulse that travels down an input Josephson transmission line and hits the storage for a short time. To measure the state of the memory: We can apply the same current pulse and measure the voltage difference across the capacitor. [This is a destructive measurement, meaning that the state is always guaranteed to end up in 1 after the measurement.] Then by integrating the voltage, we can see if the particle hopped minima or not: If we send in a pulse $I_b > I_c$  and the integrated voltage is 0, then the state was in 1 (i.e.~did not hop); if the integrated voltage is $\sim \Phi_0$, then the state was in 0 (i.e.~hopped). 
In practice this is done by coupling the storage loop to an output Josephson transmission line: A voltage spike propagates down the line and eventually gets amplified.

So in classical systems one approach to achieving both protection and control is to have large tunability of the potential (e.g.~via injected current); any natural small perturbations in memory will not change the state.

\section{Classical memory via trapped charge}

Modern laptops use the flash memory in the solid-state drive for storage of bits when the power is turned off \cite{sze2007physics, pavan1997flash, aritome_2015}. We  review the physics of these devices before turning to their superconducting analog: the (``noncompact'') Cooper pair box.

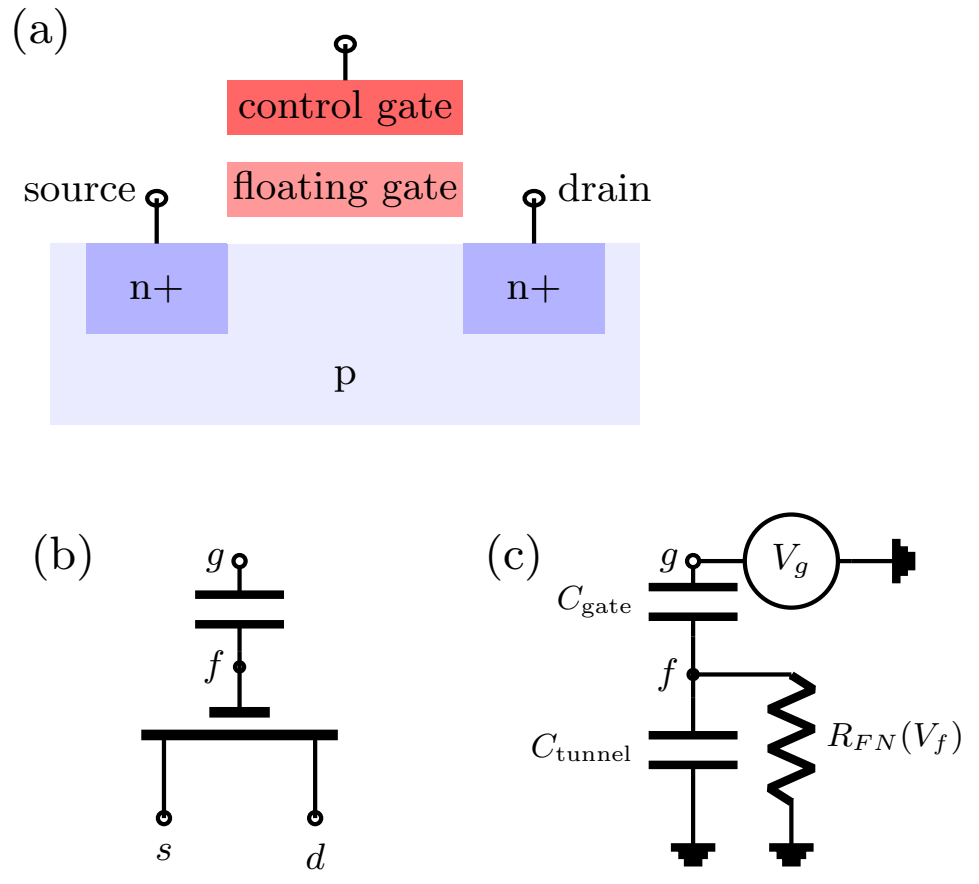
\begin{figure}
    \centering

\def\figthreefontscale{1.1} 
\def\figthreeascale{1.2} 
\begin{tikzpicture}[thick, scale=0.5, every node/.style={scale=0.7}]

\begin{scope}[shift={(0,5.5)}, scale=\figthreeascale, xscale=1.3, every node/.style={scale={0.7*\figthreeascale}}]

\fill[blue!8] (0,0) rectangle (5,2);
\node at (2.5,0.5) {p};

\fill[blue!30] (0.3,1) rectangle (1.5,2);
\node at (0.9,1.5) {n+};
\fill[blue!30] (3.5,1) rectangle (4.7,2);
\node at (4.1,1.5) {n+};

\fill[white] (1.5,2) rectangle (3.5,2.3);

\fill[red!40] (1.5,2.3) rectangle (3.5,2.9);
\node[font=\small] at (2.5,2.6) {floating gate};

\fill[white] (1.5,2.9) rectangle (3.5,3.2);

\fill[red!60] (1.5,3.2) rectangle (3.5,3.8);
\node[font=\small] at (2.5,3.5) {control gate};

\node[above] at (0.25,2.3) {source};
\draw (0.9,2) -- (0.9,2.5);
\draw (0.9,2.5) circle (0.08);

\node[above] at (4.7,2.3) {drain};
\draw (4.1,2) -- (4.1,2.5);
\draw (4.1,2.5) circle (0.08);

\draw (2.5,3.8) -- (2.5,4.2);
\draw (2.5,4.2) circle (0.08);

\node[anchor=north west, scale={1.4/\figthreeascale}] at (-0.5,4.8) {(a)};

\end{scope}

\begin{scope}[shift={(0,-0.5)}]

\node[left, scale=\figthreefontscale] at (2.5,4.2) {$g$};
\draw (2.5,4.2) circle (0.08);

\draw (2.5,4.12) to[C] (2.5,3.0);

\draw (2.5,2.8) circle (0.06);
\node[left, scale=\figthreefontscale] at (2.5,2.8) {$f$};

\draw (2.5,3.0) -- (2.5,2.2);
\draw[line width=2pt] (2.1,2.2) -- (2.9,2.2);

\draw[line width=2pt] (1.2,1.9) -- (3.8,1.9);

\draw (1.5,1.9) -- (1.5,0.8);
\draw (1.5,0.8) circle (0.08);
\node[below, scale=\figthreefontscale] at (1.5,0.65) {$s$};

\draw (3.5,1.9) -- (3.5,0.8);
\draw (3.5,0.8) circle (0.08);
\node[below, scale=\figthreefontscale] at (3.5,0.65) {$d$};

\node[anchor=north west, scale=1.4] at (-0.5,4.8) {(b)};

\end{scope}

\begin{scope}[shift={(6,-0.5)}]

\node[left, scale=\figthreefontscale] at (2.5,4.2) {$g$};
\draw (2.5,4.2) circle (0.08);

\draw (2.58,4.2) -- (3.2,4.2);
\node[draw, circle, minimum size=0.8cm, inner sep=0pt, scale=\figthreefontscale] (Vsrc) at (3.8,4.2) {$V_g$};
\draw (Vsrc.east) -- (4.6,4.2);
\node[ground, rotate=90] at (4.6,4.2) {};

\draw (2.5,4.12) to[C, l_=$C_\mathrm{gate}$] (2.5,3.2);

\draw (2.5,3.2) -- (2.5,2.4);

\draw (2.5,2.7) circle (0.06);
\node[left, scale=\figthreefontscale] at (2.5,2.7) {$f$};

\draw (2.5,2.4) to[C, l_=$C_\mathrm{tunnel}$] (2.5,1.0);
\node[ground] at (2.5,1.0) {};

\draw (2.5,2.7) -- (3.8,2.7);
\draw (3.8,2.7) to[R, l=$R_{FN}(V_f)$] (3.8,1.0);
\node[ground] at (3.8,1.0) {};

\node[anchor=north west, scale=1.4] at (-0.5,4.8) {(c)};

\end{scope}

\end{tikzpicture}

    \caption{(a) Floating-gate transistor in flash memory: 0 is represented via $Q_f\sim-3\times10^4e$ net charge on the floating gate; 1 has no charge $Q_f \sim 0$. (b) Classical circuit diagram of (a). (c) Classical circuit description of Fowler-Nordheim tunneling: A  resistor (that is exponentially sensitive to $V_f$) sits between the floating gate and ground.}
    \label{fig:floating_gate_transistor}
\end{figure}

Figure~\ref{fig:floating_gate_transistor}(a) depicts a ``floating gate transistor'':  The minimal unit needed to store a classical bit in flash memory.   The bit is stored in the floating gate: a metallic island that is surrounded by insulator.  1 is represented by no net charge on the island; 0 is represented by  $\sim30000$ net electrons on the island.   The floating gate sits between the gate voltage node  and the source-drain nodes of the transistor. [See Fig.~\ref{fig:floating_gate_transistor}(b).] The floating-gate transistor can be viewed as a regular transistor where the threshold gate voltage   depends on the  gate charge: $V_{th}(Q_{f})$. One can define: $V_{th,0} = V_{th}(Q_{f}=-30000e); V_{th,1} = V_{th}(Q_{f}=0)$.  Then in order to measure whether the floating gate is in state 0 or 1, the gate voltage is set to the middle of the two threshold voltages $V_g = (V_{th,0} + V_{th,1} )/2$, and one measures the current passing from source to drain after applying a DC voltage $V_{sd}$. A transistor  has a resistance from source to drain that is exponentially sensitive to whether the gate voltage is above/below the threshold voltage. If we assume that the floating gate is in state 1, then $V_g> V_{th,1}$ [assuming $V_{th,0}> V_{th,1}$] and current passes through from source to drain; if the state is in 0, then no current passes through.

\begin{figure}
    \centering
\begin{tikzpicture}[thick, scale=0.62, every node/.style={scale=0.85}]

\begin{scope}[shift={(0,0)}]
  \node[anchor=north west] at (-0.9,3.5) {(a)};
  \node at (2.6,3.2) {\small memory};

  \fill[blue!10] (0,-1) rectangle (2,0);
  \fill[blue!10] (3.2,-1) rectangle (5.2,0);

  \draw[very thick] (0,0) -- (2,0) -- (2,2) -- (3.2,2) -- (3.2,0) -- (5.2,0);

  \node[left] at (0,0) {$E_F$};
  \node[right] at (5.2,0) {$E_F$};

  \draw[{Stealth}-{Stealth}] (2.25,0) -- (2.25,2);
  \node[right] at (2.25,1.0) {$U_B$};

  \draw[{Stealth}-{Stealth}] (2,2.35) -- (3.2,2.35);
  \node[above] at (2.6,2.35) {$d_m$};

  \node at (1,-1.45) {\small ground};
  \node at (2.6,-2.1) {\small oxide};
  \node at (4.3,-1.45) {\small floating gate};
\end{scope}

\begin{scope}[shift={(7.2,0)}]
  \node[anchor=north west] at (-0.9,3.5) {(b)};
  \node at (2.6,3.2) {\small write};

  \fill[blue!10] (0,-1) rectangle (2,0);
  \fill[blue!10] (3.2,-4) rectangle (5.2,-3);

  \draw[very thick] (0,0) -- (2,0) -- (2,2) -- (3.2,-1) -- (3.2,-3) -- (5.2,-3);

  \node[left] at (0,0) {$E_F$};
  \node[right] at (5.2,-3) {$E_F$};

  \draw[{Stealth}-{Stealth}] (2,-0.35) -- (2.8,-0.35);
  \node[below] at (2.4,-0.4) {$d_w$};

  \draw[dashed] (4.4,0) -- (5.1,0);
  \draw[{Stealth}-{Stealth}] (4.75,0) -- (4.75,-3);
  \node[left] at (4.75,-1.5) {$eV_f$};

  \node at (1,-1.45) {\small ground};
  \node at (4.2,-4.45) {\small floating gate};
\end{scope}

\end{tikzpicture}
    \caption{ Band diagram of the floating-gate structure. (a) Memory: An electron at the Fermi level of the grounded substrate faces a rectangular barrier of height $U_B$ and width $d_m$; the transmission $\sim e^{-c d_m}$ is negligible. (b) Write: The applied voltage tilts the barrier into a triangle and pulls the bands of the floating gate down by $eV_f$. The electron now only needs to tunnel a distance $d_w < d_m$ before emerging in the oxide conduction band and drifting to the floating gate.
    }
    \label{fig:fn_band}
\end{figure}
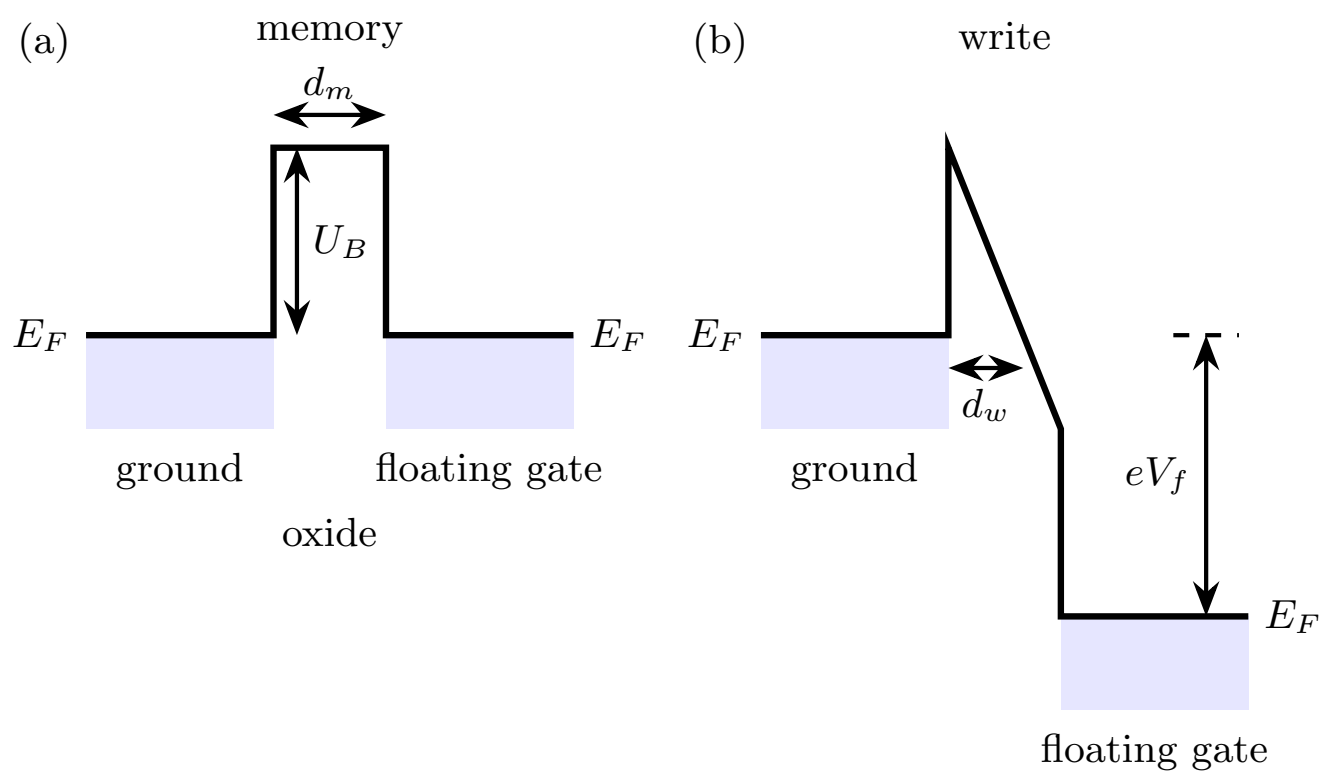

If the charge stored on the floating gate  is so stable, then how do we get it on there in the first place, i.e.~how do we write?  We apply a   voltage pulse  on the gate node that allows electrons from the ground (semiconductor) substrate to quantum tunnel across the insulator  onto the island (called Fowler-Nordheim tunneling \cite{fowler1928electron}). The applied voltage essentially reduces the tunneling barrier from ground to the floating island such that the quantum tunneling amplitude becomes significant during a write protocol. [The tunneling rate is exponential in the barrier width, which is tunable via the applied voltage; see Fig.~\ref{fig:fn_band} and Ref.~\cite{lenzlinger1969fowler} for a derivation.]
Within the formalism of classical circuit theory, one models this via a voltage-dependent resistance to ground. [See Fig.~\ref{fig:floating_gate_transistor}(c).] When $V_f=0$ the resistance is effectively infinite, and electrons are trapped on the island. The tunneling current is exponentially sensitive to $V_f$ so even pulling it up a little bit (e.g.~via a quench on $V_g$) can cause electrons to tunnel onto the island for a brief time.

\begin{figure}
    \centering
    \includegraphics[width=\columnwidth]{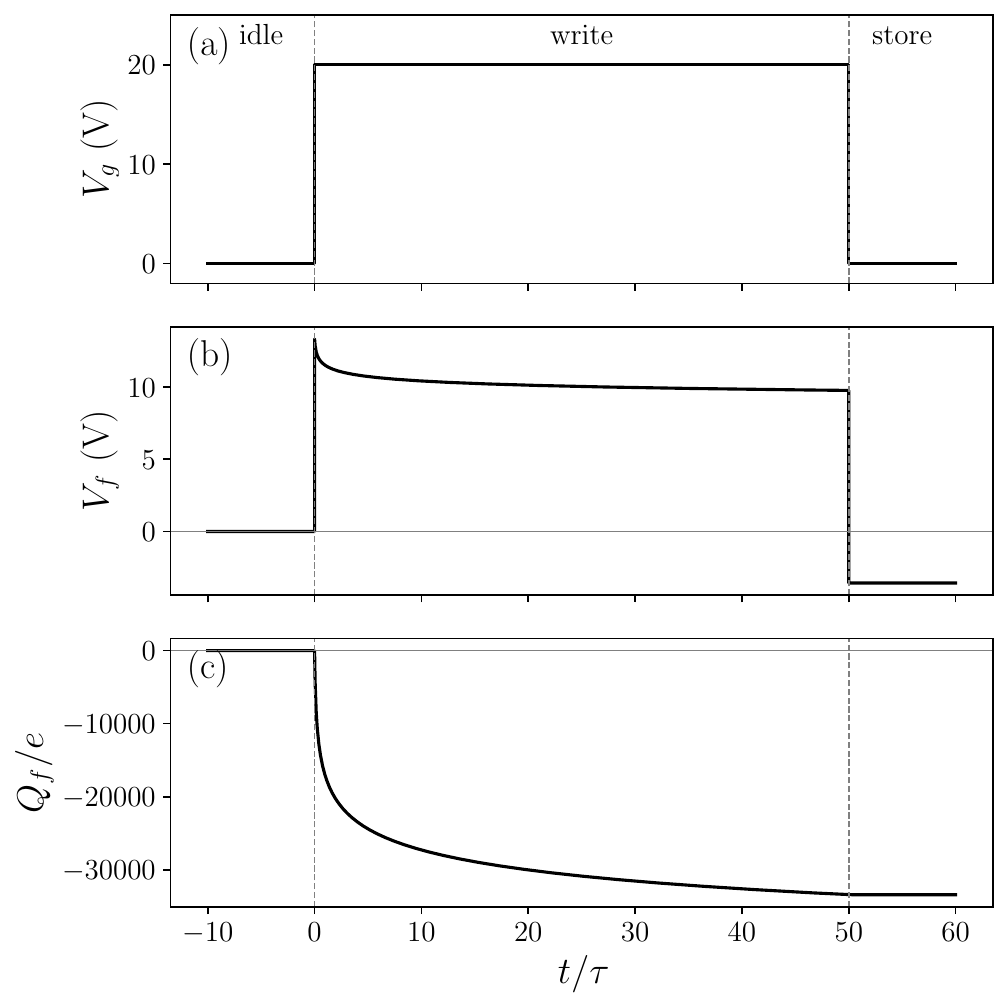}
    \caption{ A typical pulse sequence for trapping charge on the floating gate: The control gate voltage ($V_g$) rises, causing the floating gate voltage ($V_f$) to rise such that current can tunnel onto the floating gate. Parameters used:  $C_g = 1 \text{ fF},$ $ C_{\text{tun}} = 0.5 \text{ fF},$ $ V_0 = 250 \text{ V}, $ $ A = 10^{-6} \text{ A/V}^2,  $ $V_g = 20 \text{ V},$ $ \tau = 15 \text{ ms}$.}
    \label{fig:voltage_dynamics}
\end{figure}

The net charge on the floating gate is the sum of the charge on its two plates:
\begin{equation} \label{eq:float_charge}
    Q_f = C_g(V_f - V_g) + C_t V_f.
\end{equation}
Kirchhoff's current law for the evolution of $Q_f$ is:
\begin{equation}\label{eq:kirchoff_current_float}
    \frac{dQ_f}{dt} = - \frac{V_f}{R_{FN}(V_f)} = -I_{FN}(V_f).
\end{equation}
Differentiating Eq.~\eqref{eq:float_charge} (assuming $V_g$ is constant, i.e.~ignoring ramp times) and substituting into Eq.~\eqref{eq:kirchoff_current_float} leads to an equation for $V_f$:
\begin{align}
    (C_g+C_t) \frac{dV_f}{dt} &= -I_{FN}(V_f) = -A V_f |V_f| \exp\left(-\frac{V_0}{|V_f|}\right) \\
    V_f &= \frac{Q_f + C_g   V_g}{C_g+C_t}.
\end{align}
The $I_{FN}$ current is exponentially small in $V_0/|V_f|$, meaning that the current is exponentially sensitive to the voltage on the floating gate. The control gate voltage briefly raises $V_f$ such that a small amount of charge tunnels onto the floating gate. After the control voltage is turned off, that charge is effectively trapped on the floating island. Fig.~\ref{fig:voltage_dynamics} shows the typical evolution for $Q_f$ during a write protocol. To remove electrons from the island, one simply flips the sign of the control gate voltage to cause electrons to tunnel off the floating gate back to ground.

The floating gate transistor works with normal metal at room temperature. We now describe a superconducting variant of the floating gate that is capable of storing classical information in whether the floating island has a single net Cooper pair or not, i.e.~the classical-bit limit of the charge qubit. [To our best knowledge such a memory has not  been experimentally realized, likely because the quantum phase slip junction \cite{mooij2006superconducting, astafiev2012qps, kou_2025}   has been harder to reliably fabricate particularly at large phase-slip energies $E_Q$ (and charge noise is typically much larger than flux noise in superconducting platforms \cite{paladino_2014}).]  Consider the circuit in Fig.~\ref{fig:classical_memory_cartoon}(d), i.e.~a voltage source, capacitor, quantum phase slip junction, inductor, and resistor (not shown) all connected in series. A quantum phase slip junction connects two nodes via a thin superconducting wire, typically with a width that is less than the coherence length of the superconductor. This allows flux to tunnel across the junction.  This circuit is the electromagnetic dual of the flux storage loop (where  parallel $\leftrightarrow$ series, $I \leftrightarrow V$, $E_J \leftrightarrow E_Q$).   In the limit where the capacitive energies dominate the inductive energy ($E_Q,E_C \gg E_L$),  the quantum state has near-definite charge. Writing Kirchhoff's law for the voltage drop across the loop, the equation of motion for the number of Cooper pairs $n$ that have passed across each element is:
\begin{equation}
    2e V_b = \left( \frac{\hbar^2}{ E_L} \right) \ddot{n} + \left( 4e^2R \right) \dot{n} + 2\pi E_Q\sin(2\pi n) + 8 E_C ( n-1/2),
\end{equation}
where we have assumed a DC offset voltage $V_{b,0} = e/C$ [responsible for the $8E_C(-1/2)$ term] such that $V_{b, \text{tot}} = V_{b,0} + V_b$.  Again  we can  view this equation of motion as a classical particle in a potential
\begin{equation}
    U(n) = -2eV_b n - E_Q\cos(2\pi n) + 4E_C (n-1/2)^2,
\end{equation}
i.e.~a double well potential with minima near $n=0,1$,
with mass $ \hbar^2/E_L$ and damping $4e^2R$.   Having mapped the system exactly to the electromagnetic dual of the SFQ flux memory, one can now in principle use  the same approaches to  do the memory and write protocols by tuning the bias voltage.

\section{Towards protecting a qubit}

Having understood the basics of  classical memories via trapped flux and charge, we now turn to the quantum problem: Why are these circuits not good qubits and how can we design quantum circuits that protect against both bit and phase flips?

The quantum Hamiltonian corresponding to the flux storage loop circuit in Fig.~\ref{fig:classical_memory_cartoon}(c) is the flux qubit Hamiltonian:
\begin{equation}\label{eq:flux_qubit}
    H_{\text{flux}} = 4E_C n^2 + \frac{E_L}{2} (\phi-\pi + \epsilon_\phi(t))^2 - E_J \cos(\phi),
\end{equation}
where $\epsilon_\phi(t)$ is a fluctuating external flux bias through the loop. [Here we choose no DC current bias, but a $\Phi_{\text{ext}} = \Phi_0/2$ external flux bias; note that an external flux bias can always be written as an injected current, i.e.~mathematically they both couple to the $\phi$ term in the Hamiltonian.]  If we define the codewords via:  $|\bar{0}\rangle = |\phi \approx 0 \rangle; |\bar{1}\rangle =  |\phi \approx 2 \pi \rangle$,  the bit-flip time is exponentially long: $T_{bf} \sim \exp[c E_J/(k_B T)]$. However the phase-flip time is linearly sensitive to the flux noise through the loop: $T_{pf} \sim 1/ (E_L \sqrt{ \langle\epsilon_\phi^2\rangle} )$. Thus flux noise fluctuations cause pure dephasing in the model. The motivation to use superinductors in fluxonium comes directly from the prefactor: $E_L \sim 1/L$, i.e.~the coherence time scales linearly with the inductance. In App.~\ref{sec:sym_breaking}, we discuss why   a double-well potential is generically only good at protecting a classical bit.

Analogously, the noncompact Cooper pair box  (Fig.~\ref{fig:classical_memory_cartoon}(d) at half charge voltage offset) has the dual Hamiltonian \cite{osborne2024, osborne2024b}:
\begin{equation}\label{eq:noncompact_cooper}
    H_{\text{charge}} = 4E_C (n-1/2+\epsilon_n(t))^2 + \frac{E_L}{2} \phi^2 - E_Q \cos(2\pi n),
\end{equation}
where: $\epsilon_n(t)$, i.e.~is a fluctuating external charge bias on the island.  If we define the codewords to be those of the charge qubit:  $|\bar{0}\rangle = |n = 0 \rangle; |\bar{1}\rangle =  |n = 1 \rangle$ then again the bit-flip time is exponentially-long in $E_Q/(k_BT)$ while the phase-flip time is $T_{pf} \sim 1/ (E_C \sqrt{ \langle\epsilon_n^2\rangle} )$.   We note one important experimental asymmetry: In modern experiments, charge noise fluctuations are  about 3 orders of magnitude  stronger than flux noise \cite{paladino_2014}. This is one of the reasons why heavy fluxonium is generally considered to be a better qubit than the Cooper pair box.

We elaborate on why we call the model in Eq.~\eqref{eq:noncompact_cooper} the noncompact Cooper pair box. The (compact) Cooper pair box \cite{nakamura1999coherent} has the Hamiltonian:
\begin{equation} \label{eq:cooper_pair_box}
    H_{\text{cpb}}=4E_C(n-1/2+\epsilon_n(t))^2 - E_J \cos(\phi),
\end{equation}
 where typically one works in the regime $E_C \geq E_J$ (opposite of the transmon limit). The potential  is invariant under $\phi \rightarrow \phi+2\pi$, which means that its Hilbert space is that of a rotor (a compact variable). This implies \cite{pegg1989phase} that its conjugate variable is integer quantized: $n \in \mathbb{Z}$, i.e.~the number of Cooper pairs on the island is discrete. In contrast, the model in Eq.~\eqref{eq:noncompact_cooper} does not have a rotor translation symmetry and is hence noncompact (both $n,\phi$ can take an arbitrary continuous value). Nevertheless, the two models can be used to describe the extreme limit of the Cooper pair box: A floating island of superconductor  that is well separated  from ground  via an insulator.  In the compact Cooper pair box \eqref{eq:cooper_pair_box},  $E_J \sim \exp[ - d/\xi_b]$ where $d$ is the distance between the plates and $\xi_b$ is  the decay length of the pair amplitude inside the barrier (typically $\sim$ \AA); so $E_J$ is exponentially small when $d\rightarrow \infty$.  In the noncompact Cooper pair box \eqref{eq:noncompact_cooper},  $E_Q \sim \exp[ - w / \xi_q]$ where  $w$ is the width of the thin wire from the island to ground and $\xi_q$ is proportional to the superconducting coherence length; so $E_Q $ grows exponentially  as $w\rightarrow0$ in the limit of a floating superconducting island (infinitesimally thin wire). This effectively pins $n$ to integer values due to the dominant cosine potential. In both cases, one is basically left with only the $4E_C(n-1/2)^2 $ term in the Hamiltonian (and $n\in \mathbb{Z}$). In the intermediate regime of large but finite $E_Q > E_L$, we can use the WKB formula to relate parameters of the two models \cite{glazman_2002, manu2012, nguyen2025}: 
 \begin{equation}\label{eq:wkb}
     E_J \sim (4\pi^2 E_Q^3 E_L)^{1/4}\exp[-(4/\pi)\sqrt{E_Q/E_L}].
 \end{equation}
Eq.~\eqref{eq:cooper_pair_box} can be viewed as an
 effective tight-binding model  for states localized in the cosine wells of \eqref{eq:noncompact_cooper},
 with a hopping rate determined by the barrier height ($E_Q$) and kinetic energy ($E_L$).

\section{Rules of self correction vs.~protected circuits}

It is worth pausing here to make a comparison with the classical protection in the ferromagnetic Ising model:
\begin{equation}
    H_{\text{Ising}} = -J \sum_{\langle ij \rangle} Z_i Z_j.
\end{equation}
Typically in the rules of two-level system self-correcting memories \cite{brell16},  the explicit energy scales in the Hamiltonian (e.g.~$J$) are held fixed at $O(1)$ values relative to $k_B T$ and the protection comes from scaling the system size: The memory should be exponential  in $N$ not $J/(k_BT)$.
In the $N\times N$ Ising model, there is a thermal phase transition: For a finite coupling constant $J$, if $k_B T < 2.27 J$ then  $T_{bf} \sim \exp[ c N]$, i.e.~the bit-flip time is exponential in linear system size given the temperature is below some constant critical temperature 
\footnote{The exponential-in-linear-system-size bit-flip time of the thermal Ising model does not survive explicit symmetry-breaking terms (e.g.~a longitudinal field) \cite{thorngren_2026}. However, the bit-flip rate remains exponentially suppressed, with a barrier that scales as the inverse of the perturbation strength: for weak perturbations the memory time is still very long, though it no longer grows with system size. Magnetic memories survive realistic noise strengths.}.
In contrast, the protected  classical bit in the  flux qubit Hamiltonian of Eq.~\eqref{eq:flux_qubit} violates  two of the self-correcting memory rules \cite{brell16}: 1) it has an infinite-dimensional (bosonic) Hilbert space; 2) its classical protection relies on an energy scale in the Hamiltonian ($E_J$) that is large compared to $k_B T$,
i.e.~its bit-flip time scales as $T_{bf} \sim \exp[ c E_J/ (k_B T)]$.
But the rules that we impose when designing practical memories should ultimately be informed by what is physically possible: In superconducting circuits, the Hilbert space is naturally bosonic and one can make $E_J$ practically very large. Fundamentally $E_J$ is  proportional to the critical current through the junction which is exponentially sensitive to the thickness of the insulating barrier, and also scales linearly with the area of the junction (i.e.~the cross sectional area of the wire). In contrast, there is no obvious way to increase the magnetic coupling between nearest-neighbor spins ($J$) in material ferromagnets, so it makes sense to impose the constraint that $J$ cannot be made arbitrarily big.   Note that $E_J$ is the energy barrier that needs to be overcome in switching from one current configuration to another; it is not a large energy stored in the state (which has finite current corresponding to $L I = \Phi_0/2$ flux).

When designing passive memories out of superconducting circuits, a large separation of energy scales [e.g.~$E_J/(k_B T) \gg 1$] in a bosonic Hilbert space should be considered as a legitimate protection mechanism, since real classical memories rely on such a separation. 

As we discuss below, even these relaxed rules do not easily yield exponential protection against both bit and phase flips in a realistic device when all relevant noise channels are included; we examine two promising candidates and their limitations.

\section{Floating XY model}

\begin{figure}
    \centering
\begin{tikzpicture}[scale=0.55, transform shape,
  JJ/.style={draw, rectangle, minimum size=0.6cm, inner sep=0pt,
    path picture={
      \draw (path picture bounding box.south west) -- (path picture bounding box.north east);
      \draw (path picture bounding box.north west) -- (path picture bounding box.south east);
    }}
]

\def\sp{4}

\foreach \i in {0,1,2} {
  \foreach \j in {0,1,2} {
    \node[circle, fill, inner sep=2pt] (n\i\j) at (\sp*\i, \sp*\j) {};
  }
}

\foreach \i in {0,1} {
  \foreach \j in {0,1,2} {
    \pgfmathtruncatemacro{\ni}{\i+1}
    \node[JJ] (jjh\i\j) at (\sp*\i + \sp/2, \sp*\j) {};
    \draw (n\i\j) -- (jjh\i\j.west);
    \draw (jjh\i\j.east) -- (n\ni\j);
  }
}

\foreach \i in {0,1,2} {
  \foreach \j in {0,1} {
    \pgfmathtruncatemacro{\nj}{\j+1}
    \node[JJ] (jjv\i\j) at (\sp*\i, \sp*\j + \sp/2) {};
    \draw (n\i\j) -- (jjv\i\j.south);
    \draw (jjv\i\j.north) -- (n\i\nj);
  }
}

\foreach \i in {0,1,2} {
  \foreach \j in {0,1,2} {
    \coordinate (mid\i\j) at ($(\sp*\i, \sp*\j) + (0.9,-0.9)$);
    \coordinate (gnd\i\j) at ($(\sp*\i, \sp*\j) + (1.8,-1.8)$);
    \draw (n\i\j) to[C] (mid\i\j);
    \node[draw, circle, minimum size=0.5cm, inner sep=0pt] (vs\i\j) at ($(mid\i\j)!0.5!(gnd\i\j)$) {\scriptsize $V_b$};
    \draw (mid\i\j) -- (vs\i\j);
    \draw (vs\i\j) -- (gnd\i\j);
    \node[ground, rotate=-45] at (gnd\i\j) {};
  }
}

\end{tikzpicture}

    \caption{ Floating XY model: a $N\times N$ lattice of  charge islands are connected to ground only capacitively (floating), and connected to nearest neighbors via a Josephson junction (generating an XY interaction). }
    \label{fig:josephson_array}
\end{figure}
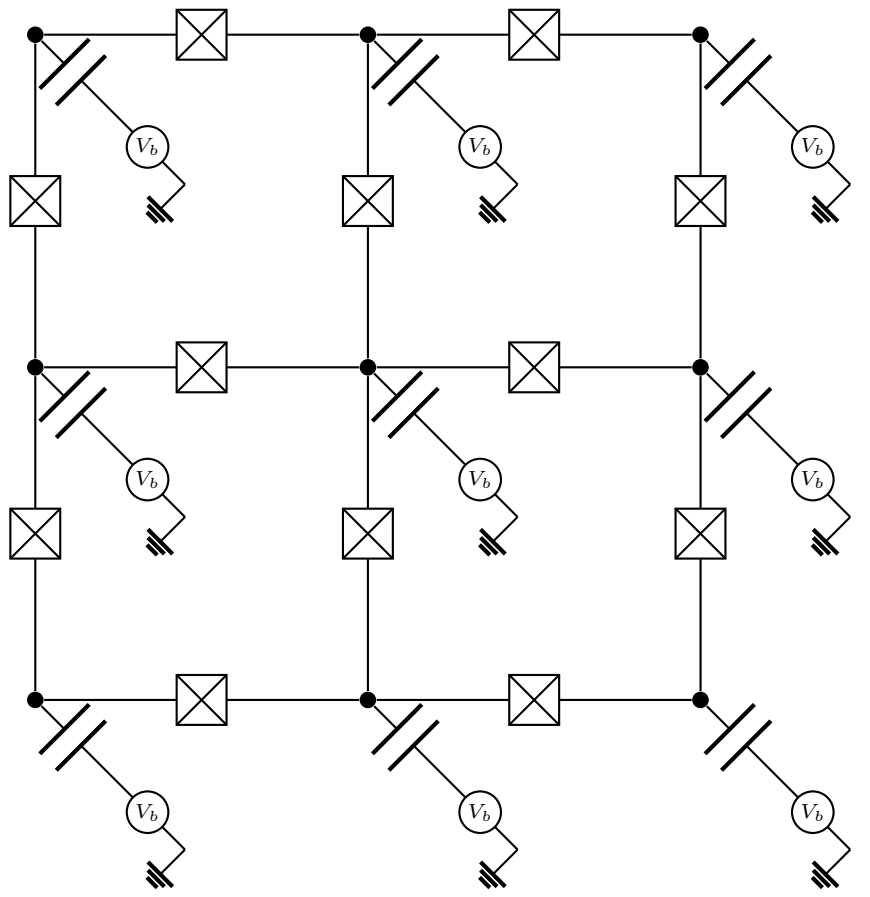

We now discuss a simple lattice  model which aims to protect a qubit by combining the (exponential) protection of a Cooper pair box with  the (polynomial) protection of  a classical XY model: The ``floating'' XY model.
Consider  an   $N \times N$ lattice Hamiltonian (see Fig.~\ref{fig:josephson_array}):
\begin{equation}
    H = 4E_{C, \text{gnd}}  \sum_j (n_j-1/2)^2 - E_J \sum_{\langle i,j \rangle} \cos(\phi_i - \phi_j),
\end{equation}
where $n_j$ is the net charge on each lattice site (charge island), $\phi_j$ is its superconducting phase, and $E_{C, \text{gnd}} = e^2/(2C_\text{gnd})$.
Each node $j$ can be viewed as a compact ($V_b=e/C_\text{gnd}$ biased) Cooper pair box \eqref{eq:cooper_pair_box} to ground with $E_{J,\text{gnd}}\rightarrow0$, or a noncompact Cooper pair box \eqref{eq:noncompact_cooper} with $E_{Q,\text{gnd}} \rightarrow \infty$ [i.e.~the limit of large insulating barrier between the superconducting island and ground]. 
The purpose of the  large insulating barrier between each island and ground is to make sure that no stray Cooper pairs (or quasiparticles) can tunnel onto the lattice from ground, i.e.~the islands are ``floating'' and the total number of Cooper pairs on the lattice is strictly conserved.  These Cooper pair boxes are then connected to nearest neighbors   via  strong Josephson junctions. We work in the limit $E_J \gg E_{C, \text{gnd}}$, i.e.~the usual transmon limit but now with $E_J$ representing nearest-neighbor couplings rather than a junction to ground.

The model has a (``strong'') $U(1)$ symmetry of total charge conservation: $S = \sum_{j} n_j; U(\theta) = \exp[-i \theta S], [H, U(\theta)] = 0$.  The ground states can spontaneously break this symmetry. We define the codewords:
\begin{align}
    |\overline{0}\rangle &= \prod_j  |\phi_j=0\rangle \\
    |\overline{1}\rangle &= \prod_j  |\phi_j=\pi \rangle \\
    |\overline{+}\rangle &= (|\overline{0}\rangle + |\overline{1}\rangle)/\sqrt{2} \nonumber \\ & \sim  \sum_{N_1+N_2 +\ldots =\text{even}} |n_1 = N_1; n_2 = N_2,\ldots \rangle \\
    |\overline{-}\rangle &= (|\overline{0}\rangle - |\overline{1}\rangle)/\sqrt{2} \nonumber \\ & \sim  \sum_{N_1+N_2 +\ldots =\text{odd}} |n_1 = N_1; n_2 = N_2,\ldots \rangle.
\end{align}
We see that the $|\overline{\pm}\rangle$ states represent a uniform superposition of charge states  where the total charge summed over  all islands is even/odd (in units of the Cooper pair charge $2e$). Normally we would have to worry about phase flips, but because Cooper pairs  cannot tunnel from ground, there is no mechanism for  $|\overline{+}\rangle$ to evolve to $|\overline{-}\rangle$ or vice-versa. More precisely, if we include a finite $E_{Q, \text{gnd}}$ between each node and ground, then (thermal) tunneling from ground is exponentially suppressed in $E_{Q,\text{gnd}}/(k_BT)$, which we take to be large. The logical phase-flip time will obey:  
\begin{align}
    T_{pf}\sim \exp[c E_{Q,\text{gnd} } / (k_B T)]/N^2
\end{align}
where the factor of $N^2$ in the denominator appears since a Cooper pair tunneling from ground onto any of the $N^2$ sites
induces a logical phase flip, so the total phase-flip rate is enhanced by a factor of $N^2$ \cite{lieu_2025}. [Quantum tunneling of Cooper pairs is instead suppressed exponentially in $\sqrt{E_{Q,\text{gnd} }/ E_{L,\text{gnd}} }$, cf.~Eq.~\eqref{eq:wkb}; single-electron (quasiparticle) tunneling is suppressed exponentially in the thickness of the insulating barrier between island and ground, which is the same geometric mechanism that gives flash memory its retention \footnote{Quasiparticles generated within the array (e.g.~by stray-photon pair breaking) conserve the total charge of the lattice and therefore commute with the logical parity $\exp[i \pi S]$; such processes can contribute to phase diffusion (bit flips) but not to phase flips, which still require electron exchange with ground and remain geometrically suppressed.}.] We assume that the floating lattice is well isolated from ground such that the exponential dominates over the polynomial factor and we do not need to worry about phase flips. We now focus  only on the bit-flip time, i.e.~the typical time it takes $|\overline{0}\rangle$ to evolve to $|\overline{1}\rangle$.

In the limit $E_J \gg E_{C, \text{gnd}}$ the model reduces exactly to the classical XY model, and we can thus simply ask what is the classical-bit thermal memory time ($T_{bf}$) of starting in one ferromagnetic state and ending up in the opposite state. [For finite $E_J/E_{C, \text{gnd}}$ ratio (and with parasitic junction capacitance) we expect the qualitative physics to be the same, i.e.~the symmetry remains broken but the ground states do not have a simple analytic form.] The XY model is an example of $U(1)$ continuous symmetry breaking (at low temperature)  and thus its classical-bit protection is expected to be significantly weaker than that of the discrete Ising model: There is no extensive energy barrier between codewords due to gapless Goldstone modes associated with the broken $U(1)$ symmetry \cite{goldstone}.

We expect local charge noise on each island (node) to be the dominant contribution to the logical bit-flip error rate. This noise causes small perturbations of the local superconducting phase on each site ($\phi_k\rightarrow \phi_k+\delta$), causing the local phases to diffuse and eventually scramble the classical information. For concreteness, we consider a  system Hamiltonian and system-bath coupling of the form:
\begin{equation}
\begin{split}
    H_S &= -E_J \sum_{\langle ij \rangle} \cos(\phi_i - \phi_j) \equiv E(\vec{\phi}), \\
    H_{SB} &= g \sum_k n_k \otimes B_k,
\end{split}
\end{equation}
with an independent thermal bath $B_k$ coupled to each site through its charge $n_k = -i\partial_{\phi_k}$. The system eigenstates are the definite-phase configurations $|\vec{\phi}\rangle$. Repeating the standard Born-Markov master equation derivation (except for the secular approximation since the spectrum of the system is gapless), one arrives at the following continuous-time master equation for the evolution of the diagonal part of the density matrix [see App.~\ref{sec:born_markov}]:
\begin{equation} \label{eq:fokker-plank}
    \frac{\partial P}{\partial t}
= \frac{\gamma(0)}{4} \sum_k \frac{\partial}{\partial \phi_k}
\left[ \frac{\partial P}{\partial \phi_k} + \beta\, F_k\, P \right]
\end{equation}
where
\begin{equation}
    P(\vec\phi, t) \equiv \langle\vec\phi\,|\rho(t)|\vec\phi\,\rangle
\end{equation}
is the classical probability of each phase configuration, and
\begin{align}
 \gamma(\omega) &\equiv \frac{g^2}{\hbar^2}\int_{-\infty}^{\infty} dt\, e^{i\omega t}\langle B_k(t)B_k(0)\rangle, \\
    F_k &\equiv \frac{\partial E}{\partial \phi_k} = E_J \sum_{\langle jk \rangle} \sin(\phi_k - \phi_j),
\end{align}
where $\gamma(\omega) \geq 0$ is the noise spectral function of the bath (units of $1/\text{time}$), whose zero-frequency value sets the overall rate.  Eq.~\eqref{eq:fokker-plank} can be viewed as diffusion (first term on the right-hand side) plus drift (second term).  It rests on two assumptions about the bath: (i) $\gamma(0)$ is finite and nonzero, i.e.~the noise power is white at low frequencies; and (ii) the bath is thermal:  $\gamma(-\omega) = e^{-\beta\hbar\omega}\gamma(\omega)$. Assumption (i) excludes, e.g., $1/f$ charge noise, which does not admit a Markovian (Lindblad) description. The Markovian thermal bath is the standard noise model under which self-correcting memories such as the 2D Ising model and 4D toric code are analyzed \cite{Alicki2010, liu2023dissipative}; we adopt it here for a fair comparison between the two settings.

To calculate the bit-flip time (at low temperature), we are interested in the time-evolution of the global phase: $\phi_g = \sum_k \phi_k /N^2$. We initialize the system in a ferromagnetic configuration $|\bar{0}\rangle$: $\phi_k=0, \forall k\implies  \phi_g=0$; evolve for a given time; then ask when we get the first exit of $\phi_g$ from the interval $(-\pi/2, +\pi/2)$.   One can show (see App.~\ref{sec:born_markov}) that the  marginal $p(\phi_g, t)$ obeys free 1D diffusion, with a diffusion constant suppressed by $N^2$: the drift term integrates to zero because $\sum_k F_k = \sum_k \partial E/\partial\phi_k \equiv 0$ ($E$ depends only on phase differences), leaving
\begin{equation} \label{eq:phig_diffusion}
    \frac{\partial p}{\partial t} = \frac{\gamma(0)}{4N^2}\, \frac{\partial^2 p}{\partial \phi_g^2}.
\end{equation}

A bit flip is the first exit of $\phi_g$ from the interval $(-\pi/2, +\pi/2)$, starting at the center. The mean exit time $\tau(x)$ from starting point $x$ satisfies the backward-Kolmogorov equation [see App.~\ref{sec:born_markov}]:
\begin{equation}
    \frac{\gamma(0)}{4N^2}\, \tau''(x) = -1,
\qquad
\tau(\pm\pi/2) = 0
\end{equation}
such that
\begin{equation}
\tau(x) = \frac{2N^2}{\gamma(0)}\left[ (\pi/2)^2 - x^2 \right],
\end{equation}
and evaluating at the center $x = 0$ gives exactly
\begin{equation} \label{eq:Tbf_fp}
    T_{ bf} = \frac{\pi^2 N^2}{2\,\gamma(0)}.
\end{equation}
Specifying to the case of Ohmic spectral density (i.e.~a coupling-weighted density of states  that scales linearly with frequency at low frequency; this is the most pessimistic bath within the Markovian class), we arrive at
\begin{equation} \label{eq:Tbf_fp2}
    T_{ bf} \sim \frac{N^2 \hbar}{\alpha k_B T}  .
\end{equation}
where we have assumed $\gamma(0) = 2\pi \alpha k_B T / \hbar$ and $\alpha$ is a dimensionless system-bath coupling strength
\footnote{For a dimensionless Ohmic coupling
$\alpha = 10^{-4}$ at $T = 20$ mK, $\gamma(0)^{-1} =
\hbar/(2\pi\alpha k_B T) \approx 0.6\ \mu\text{s}$, so
Eq.~\eqref{eq:Tbf_fp} gives $T_{bf} \approx 3\ \mu\text{s} \times N^2$,
i.e.~$\approx 30$ ms for $N = 100$. These absolute numbers are modest but our emphasis is on the scaling: $T_{bf}$ grows polynomially with lattice size at fixed temperature and coupling.}.

\begin{figure}
    \centering
    \includegraphics[width=\columnwidth]{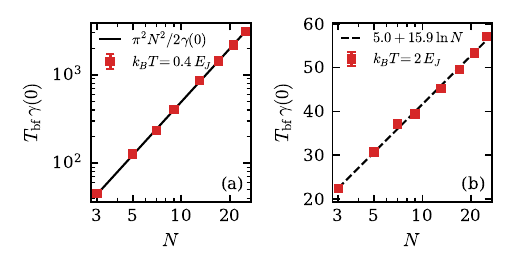}
    \caption{Bit-flip time of the floating XY model from numerical integration of the Langevin dynamics corresponding to Eq.~\eqref{eq:fokker-plank}: $\dot\phi_k = - \gamma(0)\beta F_k /4 + \sqrt{\gamma(0)/2}\,\xi_k(t)$, with $\langle \xi_k(t)\xi_l(t')\rangle = \delta_{kl}\delta(t-t')$. Each trial starts in the codeword $\phi_k = 0$, and $T_{bf}$ is the mean first exit time of the magnetization angle $\theta = \arg \sum_k e^{i\phi_k}$ from the interval $(-\pi/2, \pi/2)$; $2\times 10^3$ trials per point on an $N\times N$ periodic lattice. (a) Below the BKT transition ($k_B T = 0.4\, E_J$), the data agree with the parameter-free prediction of Eq.~\eqref{eq:Tbf_fp} (solid line) at the few-percent level. (b) Above the transition ($k_B T = 2\, E_J$), the collective $N^2$ protection is lost and the bit-flip time grows only logarithmically with system size (dashed line is a two-parameter fit).}
    \label{fig:langevin_tbf}
\end{figure}

We see that the bit-flip time is expected to grow linearly with the number of sites on the 2D lattice, and is inversely proportional to temperature and coupling strength. While this polynomial scaling is not impressive compared to the exponential scaling (in $N$) of the 2D Ising model, the polynomial protection is still  a nontrivial property of the  ordered (low-temperature) phase of the XY model:  at temperatures above the BKT transition ($k_B T \gtrsim 0.9\, E_J$), the correlation length $\xi(T)$ is finite and the array fluctuates as $\sim (N/\xi)^2$ effectively independent patches, so the collective $N^2$ suppression is lost. In that regime we expect the bit-flip time to saturate, growing at most logarithmically with system size, $T_{bf} \sim \ln N$. The $N^2$ scaling is thus a genuine property of the ordered phase, and $E_J$ enters the memory time only through the requirement $k_B T < k_B T_{\text{BKT}} \approx 0.89\, E_J$. Both behaviors are confirmed by direct simulation of the Langevin dynamics associated with Eq.~\eqref{eq:fokker-plank} [see Fig.~\ref{fig:langevin_tbf}]: below the transition the measured bit-flip time matches Eq.~\eqref{eq:Tbf_fp} with no fitting parameters, while above it $T_{bf}$ grows only logarithmically with $N$. The logarithm is the lifetime of a majority vote over independent
bits: above the transition the magnetization is a sum of
$\sim (N/\xi)^2$ independent patch phasors, so its component along
the initial direction decays as $e^{-t/\tau}$ (with $\tau$ the
$N$-independent patch decorrelation time) while the random
contribution of the patches sets an angular noise floor
$\sim \xi/N$. The magnetization angle therefore retains its initial
sign only until $e^{-t/\tau} \sim \xi/N$, i.e.
$T_{bf} \sim \tau \ln(N/\xi)$.

It is worth situating the polynomial scaling $T_{bf} \sim N^2$ within the broader landscape of passive memories. The protection here is purely kinetic: the Goldstone mode is gapless and a logical flip costs no energy at all; the suppression arises because the global phase $\phi_g$ carries an effective inertia $\propto N^2$ and must diffuse across the logical interval. This mechanism is qualitatively distinct from ``partial self-correction'' in spin models such as the 3D cubic code \cite{haah2011, bravyi2013} (see also \cite{gu2025layer, williamson2025partial}), where an energy barrier that grows logarithmically with system size, $\sim \Delta \ln N$ (with $\Delta$ the energy cost of the elementary excitations), yields an Arrhenius memory time $\sim N^{c\Delta/(k_B T)}$: a polynomial whose exponent grows upon cooling. This benefit saturates at an optimal size $N^* \sim e^{c \Delta/(k_B T)}$, beyond which the entropy of error pathways outweighs the barrier; evaluating the memory time at $N^*$ caps it at $\exp[c(\Delta/k_B T)^2]$ \cite{bravyi2013}. By contrast, the $N^2$ law has a temperature-independent exponent and does not saturate: it persists to arbitrary $N$ because the system sits in a genuine ordered phase. 
The floating XY model thus occupies a distinctive rung between unprotected qubits and genuinely self-correcting memories \cite{brown2016}: barrier-free, polynomially-protected bit flips [exponentially-protected phase flips], and scalable within an ordered phase.

We discuss a write protocol:  Given that the XY model above has strict Cooper pair number conservation (since it is floating) how do we initialize the codewords in the first place since they  are superpositions of states in different total charge sectors? Analogous to the classical memory, one could introduce a strong bias towards the $|\overline{0}\rangle = \prod_j  |\phi_j=0\rangle$ state by introducing a $ -E_{J, \text{gnd}} \sum_j \cos(\phi_j)$ term in the Hamiltonian.  Ideally this $E_{J, \text{gnd}}$ term should have exponential tunability, i.e.~it would be exponentially-small during memory (to prevent phase flips), but significant during the write protocol. One potential way to achieve this is via a tunable $E_Q$ term: A charge quantum interference device (cquid), the electromagnetic dual of the squid \cite{cquid}. Placing two equivalent quantum phase slip junctions in series results in a  voltage-tunable phase slip rate: $E_{Q}^{\text{eff}} = 2E_{Q} \cos(\pi n_g)$, where $n_g$ is the voltage bias on the node between them. [Analogous to $E_J^{\text{eff}} = 2E_J \cos(\phi_e/2)$ in a symmetric squid.]  Suppose we place a symmetric cquid between each node $j$ and ground (as well as an inductor in series). Then by tuning the offset voltage of the cquid, one would get an effective $E_J$ term that is exponentially sensitive to the effective $E_{Q,\text{gnd}}$ in the  cquid
\begin{align}
    E_{J,\text{gnd}}(n_g) \sim &(4\pi^2 (2E_{Q,\text{gnd}} \cos(\pi n_g) )^3 E_{L,\text{gnd} })^{1/4} \nonumber \\
    &\times \exp[-(4/\pi)\sqrt{2E_{Q,\text{gnd}} \cos(\pi n_g)/E_{L,\text{gnd} }}],
\end{align}
from the WKB formula \eqref{eq:wkb}.  In principle, by tuning $n_g$ from $0$ to $0.5$ one can tune the barrier height from $2E_{Q,\text{gnd}}$ to $0$. Thus by linearly varying $n_g$ one can get an exponentially-varying effective $E_{J,\text{gnd}}$: set the offset voltage to $n_g=0$ for memory, and  near $n_g\approx0.5$  to write.  Of course (as mentioned before) the quantum phase slip junction is not easy to fabricate, and there are many other questions to address before this device becomes practical such as how to: prepare arbitrary states, do readout, or do gates;  we do not comment on these here.
We note that the cquid coupler is a galvanic connection from the XY island to ground along which Cooper pairs (and quasiparticles) can diffuse, so the write circuitry might set a residual floor on the phase-flip rate.

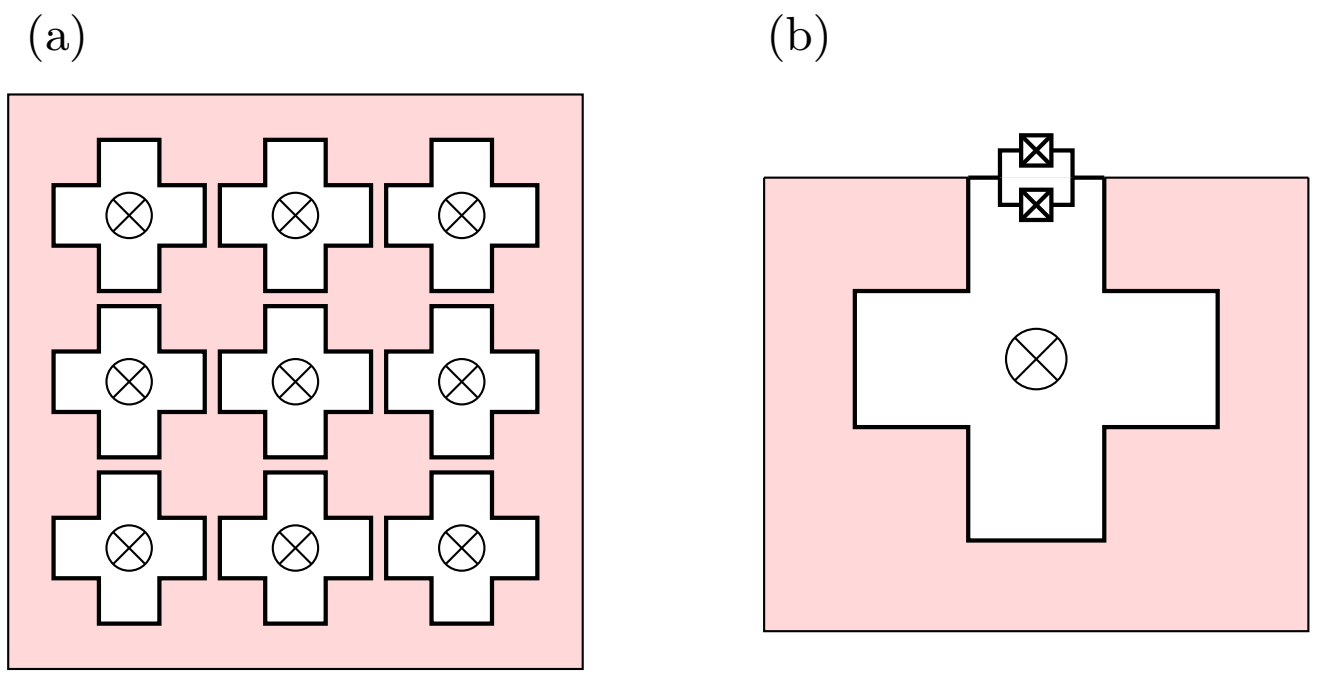
\begin{figure}
    \centering

\begin{tikzpicture}
  \def\spacing{1.1}
  \def\radius{0.5}
  \def\pad{0.3}
  
  \node[anchor=south west] at ({-\radius-\pad},{2*\spacing+\radius+\pad+0.1}) {(a)};
  
  \fill[red!15] ({-\radius-\pad},{-\radius-\pad}) rectangle ({2*\spacing+\radius+\pad},{2*\spacing+\radius+\pad});
  \draw[black] ({-\radius-\pad},{-\radius-\pad}) rectangle ({2*\spacing+\radius+\pad},{2*\spacing+\radius+\pad});
  
  \def\aw{0.2} 
  \foreach \i in {0,1,2} {
    \foreach \j in {0,1,2} {
      \begin{scope}[shift={(\i*\spacing,\j*\spacing)}]
        \fill[white] (-\aw,\radius) -- (-\aw,\aw) -- (-\radius,\aw) -- (-\radius,-\aw) -- (-\aw,-\aw) -- (-\aw,-\radius) -- (\aw,-\radius) -- (\aw,-\aw) -- (\radius,-\aw) -- (\radius,\aw) -- (\aw,\aw) -- (\aw,\radius) -- cycle;
        \draw[black, thick] (-\aw,\radius) -- (-\aw,\aw) -- (-\radius,\aw) -- (-\radius,-\aw) -- (-\aw,-\aw) -- (-\aw,-\radius) -- (\aw,-\radius) -- (\aw,-\aw) -- (\radius,-\aw) -- (\radius,\aw) -- (\aw,\aw) -- (\aw,\radius) -- cycle;
        \draw[black] (0,0) circle (0.15);
        \draw[black] (-0.105,-0.105) -- (0.105,0.105);
        \draw[black] (0.105,-0.105) -- (-0.105,0.105);
      \end{scope}
    }
  }
  
  \begin{scope}[xshift=6cm, yshift=1.25cm]
    \def\sq{1.2}
    \def\bw{0.6}
    \def\gapx{0.24}   
    \def\gapy{0.18}   
    \def\jjsize{0.10} 

    \node[anchor=south west] at ({-\sq-\bw-0.1},{1.85}) {(b)};

    \def\pw{0.45} 
    \fill[red!15] ({-\sq-\bw},{-\sq-\bw}) rectangle ({\sq+\bw},{\sq});
    \draw[black] ({-\sq-\bw},{\sq}) -- ({-\sq-\bw},{-\sq-\bw}) -- ({\sq+\bw},{-\sq-\bw}) -- ({\sq+\bw},{\sq});
    \draw[black] ({-\sq-\bw},{\sq}) -- (-\pw,\sq);
    \draw[black] (\pw,\sq) -- ({\sq+\bw},{\sq});

    \fill[white] (-\pw,\sq) -- (-\pw,\pw) -- (-\sq,\pw) -- (-\sq,-\pw) -- (-\pw,-\pw) -- (-\pw,-\sq) -- (\pw,-\sq) -- (\pw,-\pw) -- (\sq,-\pw) -- (\sq,\pw) -- (\pw,\pw) -- (\pw,\sq) -- cycle;
    \draw[black, thick] (-\pw,\sq) -- (-\pw,\pw) -- (-\sq,\pw) -- (-\sq,-\pw) -- (-\pw,-\pw) -- (-\pw,-\sq) -- (\pw,-\sq) -- (\pw,-\pw) -- (\sq,-\pw) -- (\sq,\pw) -- (\pw,\pw) -- (\pw,\sq);
    
    \draw[black] (0,0) circle (0.2);
    \draw[black] (-0.14,-0.14) -- (0.14,0.14);
    \draw[black] (0.14,-0.14) -- (-0.14,0.14);
    
    \draw[thick] (-\pw,\sq) -- ({-\gapx},{\sq});
    \draw[thick] ({\gapx},{\sq}) -- (\pw,\sq);
    
    \draw[thick] ({-\gapx},{\sq}) -- ({-\gapx},{\sq+\gapy}) -- ({-\jjsize},{\sq+\gapy});
    \draw[thick] ({\jjsize},{\sq+\gapy}) -- ({\gapx},{\sq+\gapy}) -- ({\gapx},{\sq});
    \draw[thick] ({-\gapx},{\sq}) -- ({-\gapx},{\sq-\gapy}) -- ({-\jjsize},{\sq-\gapy});
    \draw[thick] ({\jjsize},{\sq-\gapy}) -- ({\gapx},{\sq-\gapy}) -- ({\gapx},{\sq});
    
    \draw[thick] ({-\jjsize},{\sq+\gapy-\jjsize}) rectangle ({\jjsize},{\sq+\gapy+\jjsize});
    \draw[thick] ({-\jjsize},{\sq+\gapy-\jjsize}) -- ({\jjsize},{\sq+\gapy+\jjsize});
    \draw[thick] ({-\jjsize},{\sq+\gapy+\jjsize}) -- ({\jjsize},{\sq+\gapy-\jjsize});
    \draw[thick] ({-\jjsize},{\sq-\gapy-\jjsize}) rectangle ({\jjsize},{\sq-\gapy+\jjsize});
    \draw[thick] ({-\jjsize},{\sq-\gapy-\jjsize}) -- ({\jjsize},{\sq-\gapy+\jjsize});
    \draw[thick] ({-\jjsize},{\sq-\gapy+\jjsize}) -- ({\jjsize},{\sq-\gapy-\jjsize});
  \end{scope}
\end{tikzpicture}

    \caption{(a) Electromagnetic dual of the lattice in Fig.~\ref{fig:josephson_array}: A 2D sheet of superconductor (red) is punctured by an $N\times N$ lattice of plus-shaped holes (white), each of which has $\Phi_0/2$ external flux bias. Flux quanta can tunnel from one hole to its neighbor, generating  XY dynamics; the plus shape confines the quantum phase slips to the narrow superconducting bridges between facing arms of neighboring holes. (b) Zoomed-in diagram of a hole on the perimeter of the lattice [e.g.~upper middle of (a)]: A squid [not shown in (a)] separates the hole from vacuum. During memory, the squid's $E_J$ is large to prevent flux from entering the lattice; to write to the state, the $E_J$ barrier is lowered so flux can tunnel into the lattice. }
    \label{fig:xy_dual}
\end{figure}

We note that arrays of Josephson junctions like the one in Fig.~\ref{fig:josephson_array} have been used to successfully study classical XY model dynamics for several decades now \cite{resnick1981bkt, newrock2000arrays, fazio2001arrays}. In the context of this literature, the model above differs primarily in that: 1) it is important for the lattice to be ``floating'' i.e.~for Cooper pairs to be unable to tunnel from ground during memory to exponentially suppress phase flips; 2) we want to work in a large $E_J$ and small $k_B T$  regime, i.e.~deep in the ordered phase, to enhance memory effects.  Much of previous work was aimed at understanding critical properties near the transition point $E_J /h \sim k_B T /h \sim \text{10 GHz}$.

The electromagnetic dual of this model is sketched in Fig.~\ref{fig:xy_dual}(a). Here, a 2D sheet of superconductor is punctured by plus-shaped holes of insulator that can trap flux: 
\begin{equation}
    H = \frac{E_L}{2}   \sum_j (\phi_j-\pi)^2 - E_Q \sum_{\langle i,j \rangle} \cos(2\pi( n_i - n_j) ).
\end{equation}
The degrees of freedom are now how many flux quanta are trapped in each hole [$\phi_j/(2\pi) \in  \mathbb{Z}$], rather than how many Cooper pairs are on each island. An external bias of $\Phi_0/2$ threads each hole. Quantum phase slips between neighboring holes generate the XY dynamics (with a rate $\sim E_Q$). One  would like an exponentially-small quantum phase slip rate ($E_{Q, \text{gnd}} \approx 0$) from the vacuum to the lattice during memory  (i.e.~a  floating model that  conserves the total flux quanta in the lattice), but a high quantum phase slip rate from the vacuum during a write protocol. This can be achieved by placing a squid between a hole on the perimeter and the vacuum. [See Fig.~\ref{fig:xy_dual}(b).] By tuning the flux through the squid, one can exponentially control the phase slip rate across the squid (i.e.~into the hole).  Since flux noise is generally much smaller than charge noise in superconducting devices, this design might be more promising in practice.

Finally, we note that the oldest classical memory, i.e.~the abacus, derives its protection from  $U(1)$ symmetry breaking. A bead on the left of a rod represents 0; a bead on the right represents 1. The bead is a three-dimensional solid that spontaneously breaks $U(1)$ translational invariance of the rod. Naively its memory time should scale only linearly with the number of atoms in the bead: $T_{bf} \sim N_{\text{atoms}}$, but because the number of atoms is of order $N_{\text{atoms}}\sim 10^{23}$, the position of the bead remains extremely stable (it does not spontaneously slide due to the Goldstone phonon modes). (In contrast, a translationally-symmetric liquid is not a good bit.) This suggests that polynomial protection from $U(1)$ symmetry breaking can still be useful as a practical memory if the system size is large enough.

\section{Superconducting circuit with GKP ground states}

A natural question is whether a single  superconducting circuit (rather than a lattice of superconducting circuits studied in the previous section) can protect a full qubit, i.e.~the flux and charge qubit examples in Eq.~\eqref{eq:flux_qubit} and \eqref{eq:noncompact_cooper} protect a classical bit - can these circuits be promoted to protect a qubit?  Here we discuss a superconducting circuit whose degenerate ground states are GKP codewords \cite{gkp}. Recent theoretical studies have looked into various circuits that have this property \cite{Rymarz2021, le_2019,nathan_2025,cheung2026, nguyen2025}, along with a recent experimental realization \cite{nguyen2025}. We study a slight variant of the circuit described in Ref.~\cite{nguyen2025}. Our contribution is to show that the GKP codewords are symmetry-broken pointer states of  a $\mathbb{Z}_2 \times \mathbb{Z}_2$ symmetry, and to elaborate on the class of noise that is protected.

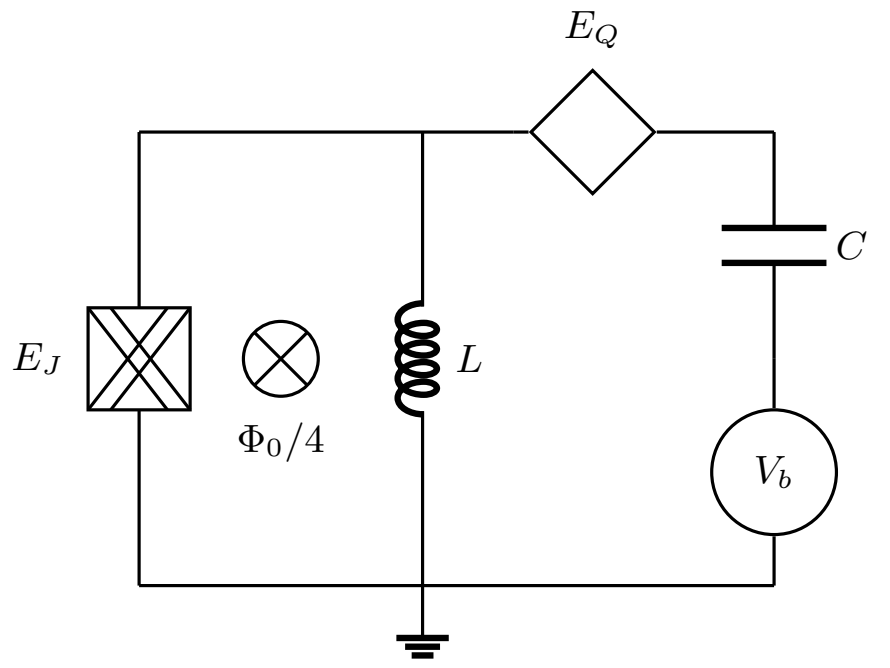
\begin{figure}
    \centering
\def\figtextscale{1.1}
\scalebox{0.75}{%
\begin{circuitikz}[thick, every node/.style={scale=\figtextscale}]

\draw (0,0) -- (0,1.55);
\draw (-0.45,1.55) rectangle (0.45,2.45);
\draw (-0.45,1.55) -- (0.25,2.45);
\draw (-0.45,2.45) -- (0.25,1.55);
\draw (-0.25,1.55) -- (0.45,2.45);
\draw (-0.25,2.45) -- (0.45,1.55);
\draw (0,2.45) -- (0,4);
\node[left, xshift=-2pt] at (-0.45,2) {$E_J$};

\draw (2.5,0) to[L] (2.5,4);
\node[right, xshift=4pt] at (2.5,2) {$L$};

\node[draw, circle, minimum size=0.6cm, inner sep=0pt,
  path picture={
    \draw (path picture bounding box.south west) -- (path picture bounding box.north east);
    \draw (path picture bounding box.north west) -- (path picture bounding box.south east);
  }] (flux) at (1.25,2) {};
\node[below, yshift=-2pt] at (flux.south) {$\Phi_0/4$};

\draw (0,4) -- (3.3,4);
\node[draw, diamond, minimum size=1cm, inner sep=0pt] (QPS) at (4.0,4) {};
\draw (3.3,4) -- (QPS.west);
\draw (QPS.east) -- (5.6,4);
\node[above, yshift=8pt] at (4.0,4.3) {$E_Q$};

\draw (5.6,4) to[C] (5.6,2);
\node[right, xshift=8pt] at (5.7,3) {$C$};
\node[draw, circle, minimum size=1cm, inner sep=0pt] (V) at (5.6,1) {$V_b$};
\draw (5.6,2) -- (V.north);
\draw (V.south) -- (5.6,0);

\draw (0,0) -- (5.6,0);
\node[ground] at (2.5,0) {};

\end{circuitikz}}
    \caption{Superconducting circuit whose degenerate ground states are GKP codewords, Eq.~\eqref{eq:gkp_ham}. A $\cos(2\phi)$ Josephson element (double-cross box), which transports \textit{pairs} of Cooper pairs, is placed in parallel with an inductor $L$, forming a storage loop threaded by a quarter flux quantum. This loop is shunted by a quantum phase slip junction (diamond) in series with a capacitor $C$; the node between them is an island biased with offset charge $e$ via $V_b$. The phase $\phi$ across the junction and the number of Cooper pairs $n$ that have crossed the phase slip junction (equal to the island charge) form the single conjugate pair of the circuit.}
    \label{fig:gkp_circuit}
\end{figure}

Consider the circuit in Fig.~\ref{fig:gkp_circuit}: a $\cos(2\phi)$ Josephson element in parallel with an inductor, shunted by a quantum phase slip junction in series with a capacitor. Applying the circuit quantization formalism of Ref.~\cite{osborne2024} (which is needed when both nonlinear charge and nonlinear flux elements are present in a circuit),
one arrives at a single-mode Hamiltonian:
\begin{equation}\label{eq:gkp_ham}
\begin{split}
    H ={}& -E_J \cos(2\phi) - E_Q \cos(2\pi n)  \\
    &+\frac{E_L}{2}\left(\phi-\frac{\pi}{2}\right)^2 + 4E_C \left(n-\frac{1}{2}\right)^2,
\end{split}
\end{equation}
where we have specified to a quarter of a flux quantum external bias through the inductive loop   and half of a Cooper pair offset charge bias on the island.  This is the simplest one-mode (self-dual) circuit model with both a  nonlinear inductor and a nonlinear capacitor that is not singular. 

The $E_J$ and $E_Q$ terms are a sum of GKP stabilizers:
\begin{equation}\label{eq:gkp_stab}
    -E_J \cos(2\phi) - E_Q \cos(2\pi n) = -E_J\, \mathrm{Re}\, S_Z - E_Q \,\mathrm{Re}\, S_X,
\end{equation}
with $S_Z = e^{2i\phi}$ and $S_X = e^{2\pi i n}$. The two stabilizers commute, and their joint $+1$ eigenspace (wavefunctions supported on $\phi \in \pi\mathbb{Z}$ and $n \in \mathbb{Z}$) is exactly two-dimensional, with logical operators $\bar{Z} = e^{i\phi}$ (transfer of a single Cooper pair) and $\bar{X} = e^{i\pi n}$ (a $\pi$ phase slip).  The doubled junction is essential: two translation operators $e^{ia\phi}$ and $e^{ibn}$ commute if and only if $ab \in 2\pi \mathbb{Z}$, and their joint $+1$ eigenspace has dimension $d = ab/(2\pi)$. A single-Cooper-pair junction [$-E_J\cos(\phi)$, i.e.~$a=1$] together with the phase slip term ($b = 2\pi$) gives $d=1$: the two cosines pin a \textit{unique} grid state with no degeneracy. This is why the flux qubit \eqref{eq:flux_qubit} and the noncompact Cooper pair box \eqref{eq:noncompact_cooper} cannot simply be merged into a qubit; doubling one of the two cosines ($a = 2$, $d=2$) is what opens a two-dimensional codespace.  The quadratic $E_L$ and $E_C$ terms (i.e.~the linear circuit elements $L$ and $C$) are needed to remove singularities in the circuit quantization,
 and provide finite energy corrections to the ideal GKP codewords. One wants to work in a regime $E_J, E_Q \gg E_L, E_C$ such that the stabilizer terms dominate the Hamiltonian.

The model has two antiunitary $\mathbb{Z}_2$ symmetries:
\begin{align}
    R_n (n-1/2) R_n^\dagger &= -(n-1/2), \qquad R_n \phi R_n^\dagger = \phi \label{eq:gkp_sym_n} \\
    R_\phi (\phi-\pi/2) R_\phi^\dagger &= -(\phi-\pi/2), \qquad R_\phi n R_\phi^\dagger = n  \label{eq:gkp_sym_phi}
\end{align}
i.e.~reflection around $n=1/2$ and reflection around $\phi = \pi/2$. [These reflections are necessarily antiunitary: a unitary reflection of a single canonical coordinate would flip the sign of the commutator $[\phi, n] = i$.] Since $R_n$ and $R_\phi$ commute as symmetry transformations, the symmetry group is $\mathbb{Z}_2 \times \mathbb{Z}_2$.

\begin{figure}
    \centering
    \includegraphics[width=0.9\columnwidth]{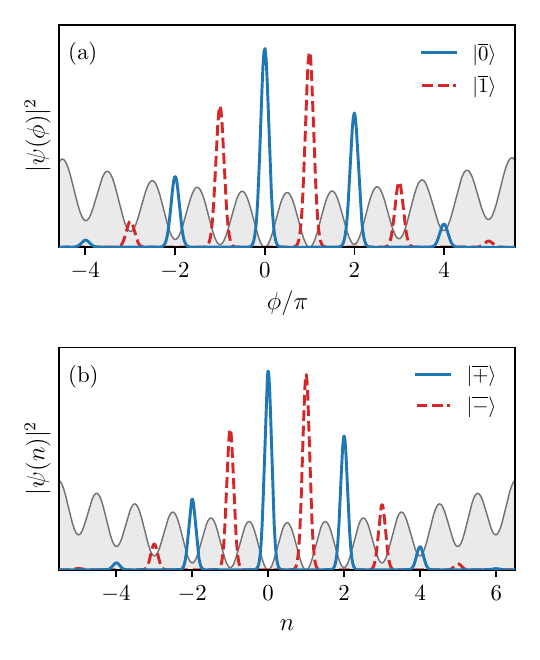}
    \caption{ Codeword profiles of the Hamiltonian \eqref{eq:gkp_ham}, from numerical diagonalization with $E_J = E_Q = 100\, E_L$ and $E_C = \pi^2 E_L/8$ (the self-dual point). A small offset is added to one of the two bias terms [a flux offset $\epsilon_\phi = 0.15$ in (a); the dual charge offset $\epsilon_n = \epsilon_\phi/\pi$ in (b)] so that the (nearly degenerate) ground doublet localizes into the symmetry-broken codewords rather than their cat combinations. (a) $|\psi(\phi)|^2$ of the two lowest eigenstates with a small flux offset: combs supported on even/odd multiples of $\pi$ ($|\overline{0}\rangle, |\overline{1}\rangle$) under a gaussian envelope. (b) $|\psi(n)|^2$ of the two lowest eigenstates with a small charge offset instead: combs supported on even/odd integers ($|\overline{\pm}\rangle$). Gray: the potential $-E_J\cos(2\phi) + (E_L/2)(\phi - \pi/2)^2$ in (a) and the dual potential $-E_Q\cos(2\pi n) + 4E_C(n-1/2)^2$ in (b). }
    \label{fig:gkp_codewords}
\end{figure}

We can define the following ideal GKP codewords:
\begin{align}
    |\overline{0}\rangle &\sim \sum_{m \in \text{even}} |\phi = \pi m \rangle \label{eq:gkp_codeword0} \\
    |\overline{1}\rangle &\sim \sum_{m \in \text{odd}} |\phi = \pi m \rangle  \\
    |\overline{+}\rangle &\sim |\overline{0}\rangle + |\overline{1}\rangle \sim \sum_{m \in \text{even}} |n  =  m \rangle \\
    |\overline{-}\rangle &\sim |\overline{0}\rangle - |\overline{1}\rangle \sim \sum_{m \in \text{odd}} |n  =  m \rangle \label{eq:gkp_codewordminus}
\end{align}
These codewords have the interesting property that they each spontaneously break one of the two antiunitary  $\mathbb{Z}_2$ symmetries of the Hamiltonian: $R_\phi |\overline{0} \rangle \sim |\overline{1} \rangle;  R_n |\overline{0} \rangle \sim |\overline{0} \rangle$, i.e.~the 0/1 codewords spontaneously break $R_\phi$ but are symmetric under $R_n$. Similarly: $R_n |\overline{+} \rangle \sim |\overline{-} \rangle;  R_\phi |\overline{+} \rangle \sim |\overline{+} \rangle$, i.e.~the $\pm$ codewords spontaneously break $R_n$ but are symmetric under $R_\phi$. Thus all pointer states of the codespace are symmetry broken states (and hence expected to be stable to perturbations, in the sense of App.~\ref{sec:sym_breaking}).
 The true ground states of the Hamiltonian \eqref{eq:gkp_ham} have a gaussian envelope  but still maintain the symmetry broken pattern. In Fig.~\ref{fig:gkp_codewords} we plot the profiles of the codewords in $\phi$ space and $n$ space [along with the potential].
 
This symmetry structure gives an alternative way to see why the GKP
circuit hosts a qubit where the circuits of
Eqs.~\eqref{eq:flux_qubit} and \eqref{eq:noncompact_cooper} host only a
classical bit: the flux qubit possesses the same two antiunitary
$\mathbb{Z}_2$ symmetries [with $R_n$ reduced to ordinary complex
conjugation, $n \rightarrow -n$, in the absence of a charge offset], but
only $R_\phi$ is spontaneously broken, since the potential is a
double well in $\phi$ while the dual ($n$) direction has a single
minimum; the noncompact Cooper pair box breaks only $R_n$, for the dual
reason. The doubled cosines of Eq.~\eqref{eq:gkp_ham} furnish degenerate
wells in both quadratures, rendering both symmetries
spontaneously breakable at once.

\begin{figure}
    \centering
    \includegraphics[width=\columnwidth]{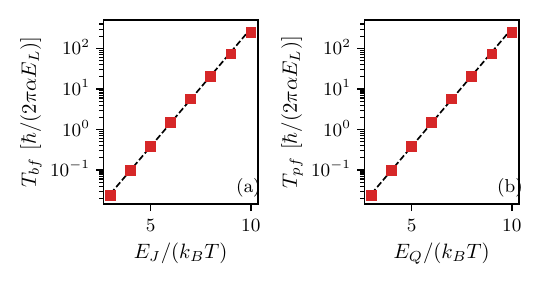}
    \caption{ Bit-flip (a) and phase-flip (b) times of the GKP circuit \eqref{eq:gkp_ham} under Ohmic charge and flux noise, from the secular Born-Markov master equation over the exact circuit eigenstates. Parameters are chosen at the self-dual point ($E_J = E_Q = 100\,E_L$, $E_C = \pi^2 E_L/8$, charge noise $\pi$ times stronger than flux noise (sweeping $T$); same Hamiltonian parameters and bias offsets as Fig.~\ref{fig:gkp_codewords}, plus a much weaker secondary offset in the dual quadrature [$\epsilon_n = 0.02/\pi$ in (a), $\epsilon_\phi = 0.02$ in (b)] that lifts accidental mirror-well degeneracies of the unbiased quadrature), so the two panels are identical by construction; times in units of $\hbar/(2\pi\alpha E_L)$; all results are converged with respect to grid size and the number of eigenstates retained. Dashed lines: Arrhenius fits $T_{bf/pf} \sim \exp[c E/(k_B T)]$ over the window $4 \leq E/(k_BT) \leq 8$. }
    \label{fig:gkp_tbf}
\end{figure}

To make our analysis concrete, we consider the following simulation: charge and flux noise couple to the system's $n$ and $\phi$ operators respectively, each to an independent Ohmic bath with the same spectral function $\gamma(\omega) = 2\pi\alpha\, \omega/(1 - e^{-\beta\hbar\omega})$ used in the XY-model analysis [so $\gamma(0) = 2\pi\alpha k_B T/\hbar$]. We work at the self-dual point of the model: the duality map is $\phi \leftrightarrow \pi n$ so the Hamiltonian \eqref{eq:gkp_ham} is self-dual for $E_Q = E_J$ and $E_C = \pi^2 E_L/8$, and the noise is self-dual when the charge coupling is $\pi$ times stronger than the flux coupling. With these choices, $T_{pf} = T_{bf}$ identically by construction. We diagonalize Eq.~\eqref{eq:gkp_ham} numerically, decompose the coupling operators in the energy eigenbasis, and evolve the populations with the standard secular Born-Markov (Pauli) master equation with golden-rule rates $W_{i\rightarrow f} = |\langle f|\phi|i\rangle|^2 \gamma(\omega_{if}) + \pi^2 |\langle f|n|i\rangle|^2 \gamma(\omega_{if})$. As in Fig.~\ref{fig:gkp_codewords}, a small bias offset localizes the eigenbasis into the comb states, so that the codeword populations are directly accessible. To measure the bit-flip time, we prepare the state in $|\overline{0} \rangle$, evolve, and record the probability $P_{\rm err}(t)$ of residing in an odd well of the $\phi$ potential [classified via $\langle \cos\phi\rangle$]; the decay fits $P_{\rm err}(t) = \frac{1}{2}(1 - e^{-t/T_{bf}})$ to better than $10^{-3}$, with $T_{bf}$ equal to the inverse of the slowest relaxation eigenvalue of the rate matrix. We find an Arrhenius law $T_{bf}\sim \exp[c E_J / (k_B T)]$. The Arrhenius form is robust at any fixed offset.
The dual measurement (charge offset, wells classified via $\langle\cos \pi n\rangle$) gives the identical $T_{pf}\sim \exp[c E_Q / (k_B T)]$.  The numerical simulations are shown in Fig.~\ref{fig:gkp_tbf}.  This agrees with recent theoretical results \cite{cheung2026} that study a generic GKP Hamiltonian coupled to a thermal bath via the $\phi, n$ operators and predict the same error scaling. 

We have  found that thermal  noise that couples to the system's $\phi, n$ operators is exponentially suppressed in the  energy scales $E_J$ and $E_Q$; is there any caveat to realistic  protection in the superconducting circuit?  The one ingredient that is  suspect in the Hamiltonian \eqref{eq:gkp_ham} is the $\cos(2\phi)$ term, i.e.~a Josephson junction that primarily allows \textit{pairs} of Cooper pairs to tunnel through the junction.  In any real implementation of a $\cos(2\phi)$ junction there is typically a residual $\cos(\phi)$ term in the Hamiltonian, i.e.~it is difficult to completely eliminate the tunneling of single Cooper pairs across a barrier. For example,  the simplest approach to obtaining an effective $\cos(2\phi)$  term is via a ``kite'' (rhombus) of Josephson junctions with half of a flux quantum external flux bias through the kite \cite{Smith2020, ioffe_2012}: destructive interference  realizes an effective $\cos(2\phi)$ junction between opposite nodes, but any flux noise in this external bias causes a fluctuating residual $\cos(\phi)$ term. Thus  in   superconducting  realizations of $\cos(2\phi)$ junctions  in general one needs to worry about  Lindblad jump terms of the form $L = \sqrt{\gamma_{\cos(\phi)}} \cos(\phi)$ due to any time-dependent $\cos(\phi)$ fluctuations. Such a term immediately dephases the qubit (i.e.~the logical dephasing rate is proportional to $\gamma_{\cos(\phi)}$). 
A  $L \sim \cos(\phi)$ jump immediately swaps the $\pm$ states: $ L |\pm \rangle \sim | \mp \rangle $, making the code ``distance-1'' in this error; $\cos(\phi)$ acts exactly as the logical $\bar{Z}$ and is therefore invisible to the stabilizers and not suppressed by the energy gap.
Note the asymmetry between the two quadratures: on the flux side, the dangerous $\pi$ phase slip $e^{i\pi n} = \bar{X}$ does not exist as a physical process (phase slips are quantized in units of $2\pi$), whereas on the charge side the dangerous single-Cooper-pair transfer $e^{i\phi} = \bar{Z}$ is a generic physical process that must be engineered away, e.g.~by interference \cite{larsen2020, Smith2020}.  The GKP qubit above thus has the property that it is exponentially protected against noise that is  coupled to  $n,\phi$,
but physical realizations  will likely come with noise terms that couple to $\cos(\phi)$
which are not protected and will set the floor of the dephasing rate.

Finally, we comment on the role of parasitic terms which we have
neglected so far: any physical Josephson element comes with a parallel
capacitance $C_J$, and any phase slip junction with a series (kinetic)
inductance $L_Q$. Including these promotes the circuit of
Fig.~\ref{fig:gkp_circuit} from one mode to three. The two additional modes
are the plasma mode of the junction, at frequency
$\sim\sqrt{E_{C_J} E_J}$, and its dual, the charge plasma mode
of the phase slip junction at $\sim\sqrt{E_{L_Q} E_Q}$; both are
pushed to high frequency as the parasitics become small. Crucially, the parasitic modes enter as harmonic oscillators coupled to the system's
$\phi$ and $n$ operators.
Absorbing the parasitic modes into the bath therefore merely yields a
structured spectral density for noise, and
the exponential protection is expected to survive.

\section{Is noise local in phase space?}

In self-correcting memories of two-level systems (e.g.~the 2D Ising model or the 4D toric code) one typically has codewords that are well separated in real space, with a noise model that is assumed to be spatially local and composed of low-weight operators. For example, the jump operators of the thermalizing Ising Lindbladian  are low-weight Paulis: $L_j^\infty \sim X_j$ and $ L_j^0 \sim X_j (1-Z_j Z_{j+1})$ for the infinite-temperature and zero-temperature jumps at site $j$ (finite temperature has both terms) \cite{liu2023dissipative}. Indeed if one were to assume jump operators that directly connected codewords (e.g.~$L\sim \prod_j X_j $ in the Ising model) then no model would  be  self correcting, since a single application of this noise would cause a logical error. Thankfully it appears that nature respects this local, low-weight property since ferromagnetic memories exist. 

For the bosonic Hamiltonians of superconducting circuits studied in this work [Eqs.~\eqref{eq:flux_qubit}, \eqref{eq:noncompact_cooper}, \eqref{eq:gkp_ham}] it is tempting to come up with a similar principle: Codewords are separated in phase space and noise terms only act locally in phase space - but is this true?   Noise that couples to the system's charge and flux operators respects this property: $\phi, n$  generate infinitesimal displacements of a coherent state in phase space.  The definite flux states in SFQ: $|\overline{0}\rangle = |\phi \approx 0 \rangle; |\overline{1}\rangle = |\phi \approx  2\pi\rangle$ are squeezed states, and local kicks via $\phi, n$ cannot move  one codeword to another without encountering an $E_J$-barrier at $\phi= \pi $ \cite{lieu_2025}.

However, we have seen (from the previous section) that in some cases jump operators of the form: $L\sim \cos(\phi)$ can also appear in the Lindbladian [e.g.~whenever a fluctuating flux threads a loop composed only of Josephson elements], and these act nonlocally on phase space, e.g.~$e^{i \phi} |n=0 \rangle = |n=1 \rangle$. In some cases (like the GKP model above) they directly move one codeword to its opposite ($+$ to $-$).  This implies that certain models naturally exhibit noise that performs finite translations in phase space.  The more general principle seems to be that  realistic noise consists of (i) infinitesimal displacements in phase space, and (ii) single-quantum tunneling events: one electron
($e^{i\phi/2}$), one Cooper pair ($e^{i\phi}$), or one fluxon
($e^{2\pi i n}$), through pathways that physically exist in the
hardware, with amplitudes that are exponentially tunable via geometry (e.g.~making an insulator thicker, or a superconducting wire thicker).  The flux qubit's classical memory is stable in the thickness of the superconducting wire (the rate of $L\sim \cos(2\pi n)$ is exponentially small in the thickness), and the charge qubit's classical memory is stable in the thickness of its insulating barrier. An ideal  passive  memory should  thus  have an answer to  noise terms that can couple to any of these system operators.

\section{Outlook}

In this work we have reviewed  two ways that real computers  store and write classical information that cannot be described by a lattice of two-level systems: by trapping charge on a conducting island surrounded by an insulator (flash memory), and by trapping flux in the middle of a superconducting loop (SFQ).  Like the Ising model, these memories are: physically realizable, thermally stable, and have a simple Hamiltonian description (via electromagnetic circuit theory). We have  explained the generic mechanisms that protect these bits: 1) These circuits rely on a large  potential barrier between codewords; the exponential protection is  due to a large separation in energy scales, e.g. $E_J/(k_BT)\sim 350$ in SFQ, rather than scaling a lattice to the thermodynamic limit. 2) In order to change the state of the system (i.e.~to write to it) we need to significantly alter the potential, e.g.~by applying a current bias or voltage bias to remove/reduce the barrier. 


While the ``lattice of two-level systems'' approach to passive error correction is general enough that it can be used to model many different physical platforms (e.g.~arrays of superconducting qubits, atoms/ions, materials, etc.), this description can sometimes miss important physics associated with specific hardware platforms.  For example, one needs to go beyond the two-level system approximation to understand why the flux qubit  can have an exponentially-small bit-flip rate: the flux qubit \eqref{eq:flux_qubit} has a bosonic Hilbert space, with codewords that are well separated in phase space and noise that is local \cite{lieu_2025}.

Looking towards a quantum memory, we discussed a ``floating XY model'' constructed out of an $N\times N$ lattice of floating charge islands connected to neighbors via Josephson junctions.  The  phase flip time is exponential in the thickness of the insulating barrier between the islands and ground  ($E_{Q,\text{gnd}} /(k_B T)\rightarrow \infty$); the bit flip time  scales as $T_{bf} \sim  N^2 / (k_BT) $, i.e.~only polynomially in system size but even this is  nontrivial, arising from  a  $U(1)$ symmetry-broken phase. We also discussed a self-dual circuit whose ground states are GKP codewords, showing that it can protect
against noise that is local in phase space but that realistic circuit implementations also host nonlocal noise that may limit performance. The search for a superconducting qubit memory device that is exponentially protected against all relevant noise channels remains open.  In this work we have focused on memory, but there are many questions to address before such devices become practical such as how to: prepare arbitrary states, do readout, and do gates; future work should investigate these directions. Our work aims  to elucidate the mechanisms of protection for  superconducting circuits,  with the goal of  expanding the search space of models  that can host a genuine passive quantum memory  beyond a lattice of two-level systems.

\begin{acknowledgments}
I thank Connor Hann, Yoni Schattner, Kevin Crowley, and Long Nguyen for inspiring discussions. Claude  assisted with mathematical derivations, numerical analysis, and manuscript preparation. I  take responsibility for the manuscript.
\end{acknowledgments}

\appendix

\section{$\mathbb{Z}_2$ spontaneous symmetry breaking} \label{sec:sym_breaking}

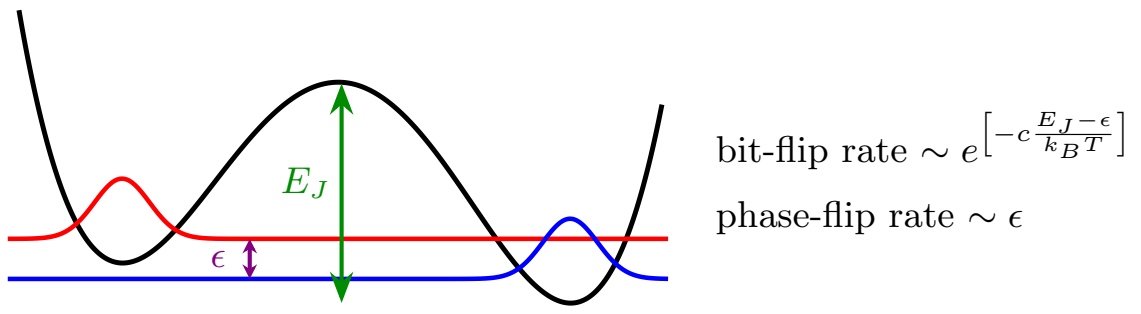
\begin{figure}
    \centering
\scalebox{0.78}{%
\begin{tikzpicture}[xscale=1.9, yscale=1.7]
  \draw[black, line width=1.2pt, domain=-1.45:1.42, samples=300, smooth]
    plot (\x, {(\x*\x-1)^2 - 0.1*\x});
  \draw[red, line width=1.1pt, domain=-1.5:1.45, samples=300, smooth]
    plot (\x, {0.22 + 0.3*exp(-((\x+0.99)^2)/0.03)});
  \draw[blue, line width=1.1pt, domain=-1.5:1.45, samples=300, smooth]
    plot (\x, {0.02 + 0.3*exp(-((\x-1.01)^2)/0.03)});
  \draw[green!55!black, {Stealth}-{Stealth}, line width=1pt] (-0.01,-0.10) -- (-0.01,0.995);
  \node[green!55!black] at (-0.17,0.5) {$E_J$};
  \draw[violet, {Stealth[length=3.5pt,width=3.5pt]}-{Stealth[length=3.5pt,width=3.5pt]}, line width=1pt] (-0.42,0.02) -- (-0.42,0.22);
  \node[violet] at (-0.56,0.12) {$\epsilon$};
  \node[anchor=west, align=left] at (1.6,0.55)
    {bit-flip rate $\sim e^{\left[-c \frac{E_J-\epsilon}{k_B T}\right]}$\\[5pt]
     phase-flip rate $\sim \epsilon$};
\end{tikzpicture}}
    \caption{  Effect of a randomly fluctuating symmetry-breaking perturbation $\epsilon(t)$ in the flux qubit (i.e.~external flux deviation  away from $\Phi_0/2$, which tilts the double well): The phase-flip rate is directly proportional to the noise magnitude. The noise enters the bit-flip rate only inside the exponent, via a slight reduction of the barrier, so bit flips remain exponentially suppressed.  The symmetry-broken states are much more robust to explicit symmetry-breaking perturbations than the cat states. }
    \label{fig:double_well}
\end{figure}

The classical memories above (and indeed the ferromagnetic Ising model) can be understood in the language of $\mathbb{Z}_2$ symmetry breaking: The classically-stable states  must spontaneously choose to sit in one of the wells at the bottom of a double well potential.
It appears that robust classical bits are associated with  $\mathbb{Z}_2$ spontaneous symmetry breaking.

The statement that ``$\mathbb{Z}_2$ spontaneous symmetry breaking protects a classical bit'' is less trivial than it seems: In quantum error correction  we typically say that   any protection that relies on  an exact symmetry is fragile (because generic errors violate all symmetries), and thus we must rely on some form of topological order for robust protection. However, while spontaneous symmetry breaking does indeed rely on the presence of a symmetry in the Hamiltonian, the resulting symmetry-broken ground states (or steady states) are very robust even to explicit symmetry-breaking perturbations. If the symmetry is violated by a term of order $\epsilon$, the logical bit-flip rate  is not directly proportional to this perturbation. For example, consider the flux qubit's  double-well potential with a bias by an amount $\epsilon$ toward the right well (e.g.~due to stray flux through the loop): The Arrhenius law  implies that it still takes an exponentially-long time [in $ (E_J - \epsilon)/ (k_B T)$] for a state localized in the left well to tunnel across since thermal excitations must kick it above the barrier (i.e.~the bit-flip rate is not linear in $\epsilon$). On the other hand, the dephasing rate of a cat superposition of symmetry-broken states $(|l\rangle + |r\rangle)/\sqrt{2}$ is linearly sensitive to $\epsilon$ since such a term induces an energy splitting that immediately leads to an unwanted relative phase between states in the left and right well: $( e^{i  \epsilon t / \hbar} |l\rangle +  |r\rangle)/\sqrt{2}$.  Systems that exhibit $\mathbb{Z}_2$ spontaneous symmetry breaking thus have a generic feature whereby explicit symmetry-breaking terms in the Hamiltonian immediately cause the phase-flip rate to increase, while the bit-flip rate is much less sensitive.  [The qubit protected by $\mathbb{Z}_2$ strong symmetry breaking is fragile to explicit symmetry breaking perturbations; the classical bit protected by $\mathbb{Z}_2$ weak symmetry breaking is robust \cite{lieu2020}.] The colloquial way of saying this is that symmetry-broken states are stable while cat superpositions are not, which is
why we observe spontaneous symmetry breaking in the real world even though symmetries are never exactly respected:  States with definite flux are stable  even in the presence of small flux perturbations
 that bias the system one way or another, but their cat superpositions are not.

\section{Local charge noise $\Rightarrow$ Fokker--Planck dynamics} \label{sec:born_markov}

This appendix derives Eqs.~\eqref{eq:fokker-plank}, \eqref{eq:phig_diffusion}, and \eqref{eq:Tbf_fp} of the main text.

\emph{Derivation of Eq.~\eqref{eq:fokker-plank}.} Second-order Born--Markov theory, before any secular approximation, yields a master equation written in terms of the eigenoperators of the system Hamiltonian \cite{breuer2002}:
\begin{equation}\label{eq:presecular}
\begin{split}
    \frac{d\rho}{dt} = \sum_k \sum_{\omega, \omega'} \gamma_k(\omega, \omega') \Big[ &A_k(\omega)\, \rho\, A_k^\dagger(\omega') \\
    &- \tfrac{1}{2} \big\{ A_k^\dagger(\omega') A_k(\omega),\, \rho \big\} \Big],
\end{split}
\end{equation}
\begin{equation}\label{eq:eigenop}
    A_k(\omega) \equiv \sum_{E} \Pi(E + \hbar\omega)\, n_k\, \Pi(E),
\end{equation}
where $\Pi(E)$ projects onto system eigenstates of energy $E$, such that $A_k(\omega)$ collects every matrix element of $n_k$ that raises the system energy by exactly $\hbar\omega$. The coefficient $\gamma_k(\omega,\omega')$ reduces to the spectral function of the main text on the diagonal, $\gamma_k(\omega,\omega) = \gamma(\omega)$ [the imaginary (Lamb-shift) parts cancel there]. The usual secular approximation would discard all $\omega \neq \omega'$ terms, but this requires Bohr-frequency spacings that are large compared to the dissipative rates, which fails for the present gapless spectrum; we keep all terms.

In the classical limit ($E_{C, \text{gnd}} \rightarrow 0$) the system eigenstates are the definite-phase configurations $|\vec{\phi}\rangle$ with energies $E(\vec{\phi})$, and the coupling operator has matrix elements $\langle \vec{\phi}\,'|n_k|\vec{\phi}\rangle = -i\,\delta'(\phi_k' - \phi_k)\prod_{j \neq k}\delta(\phi_j' - \phi_j)$. To regularize the $\delta'$, place each phase on a ring grid of spacing $a$, taken to zero at the end:
\begin{equation}\label{eq:nk_discrete}
    \langle \vec{\phi} \pm a\hat{e}_k | n_k | \vec{\phi} \rangle = \pm \frac{1}{2ia},
\end{equation}
where $\vec{\phi} \pm a\hat{e}_k$ shifts $\phi_k$ by $\pm a$ with all other sites unchanged. Thus $n_k$ consists of elementary hops $|\vec{\phi} \pm a\hat{e}_k\rangle\langle\vec{\phi}|$, with energy change $\Delta E_{k,\pm} = \pm a F_k + O(a^2)$. Each hop belongs to exactly one eigenoperator \eqref{eq:eigenop}, namely the one whose frequency $\hbar\omega$ matches its energy change, and completeness reassembles them: $\sum_\omega A_k(\omega) = n_k$.

Now project Eq.~\eqref{eq:presecular} onto the diagonal, $P(\vec{\phi}) = \langle\vec{\phi}|\rho|\vec{\phi}\rangle$, taking $\rho$ diagonal. In the first (sandwich) term, $A_k(\omega)$ and $A_k(\omega')$ must then contain the \emph{same} elementary hop $\vec{\chi} \rightarrow \vec{\phi}$; since each hop lives in a single eigenoperator, only the terms with $\omega = \omega'$ contribute. The same holds for the anticommutator term. The $\omega \neq \omega'$ terms require off-diagonal elements of $\rho$: these vanish in the (classical, diagonal) initial state and are regenerated by populations only at $O(g^2)$, so their feedback into $P$ is beyond the accuracy of the Born approximation. The projection therefore closes, yielding a Pauli master equation with golden-rule rates $W_{i\rightarrow f} = |\langle f|n_k|i\rangle|^2\, \gamma\big((E_i - E_f)/\hbar\big)$:
\begin{equation}\label{eq:pauli}
\begin{split}
    \frac{dP(\vec{\phi})}{dt} = \sum_k \sum_{\sigma=\pm} \Big[ &W_{k,-\sigma}(\vec{\phi} + \sigma a\hat{e}_k)\, P(\vec{\phi} + \sigma a\hat{e}_k) \\
    &- W_{k,\sigma}(\vec{\phi})\, P(\vec{\phi}) \Big],
\end{split}
\end{equation}
\begin{equation}\label{eq:hop_rates}
    W_{k,\pm}(\vec{\phi}) = \frac{1}{4a^2}\, \gamma(\mp a F_k/\hbar),
\end{equation}
i.e.~the squared matrix element \eqref{eq:nk_discrete} times the spectral function evaluated at \emph{minus} the system energy change, since an uphill move must extract energy from the bath. The KMS condition $\gamma(-\omega) = e^{-\beta\hbar\omega}\gamma(\omega)$ gives detailed balance, $W_{\text{up}}/W_{\text{down}} = e^{-\beta\Delta E}$.

For small $a$, Taylor-expand the gain term of \eqref{eq:pauli} about $\vec{\phi}$; the $m=0$ terms cancel the loss terms, and collecting the rest gives the Kramers--Moyal expansion
\begin{equation}\label{eq:kramers_moyal}
\begin{split}
    \frac{\partial P}{\partial t} &= \sum_k \sum_{m \geq 1} \frac{(-1)^m}{m!}\, \frac{\partial^m}{\partial \phi_k^m} \Big[ \alpha_m^{(k)}(\vec{\phi})\, P \Big], \\
    \alpha_m^{(k)} &\equiv \sum_{\sigma = \pm} (\sigma a)^m\, W_{k,\sigma},
\end{split}
\end{equation}
still exactly equivalent to \eqref{eq:pauli}; the jump moments $\alpha_m^{(k)}$ are the mean $m$-th powers of the displacement of $\phi_k$ accumulated per unit time. Assuming $0 < \gamma(0) < \infty$, KMS fixes the slope at zero frequency, $\gamma'(0) = \tfrac{1}{2}\beta\hbar\,\gamma(0)$, so $\gamma(\mp aF_k/\hbar) = \gamma(0)[1 \mp \beta a F_k/2 + O(a^2)]$ and
\begin{equation}
    \alpha_1^{(k)} = -\frac{\gamma(0)}{4}\,\beta F_k, \qquad
    \alpha_2^{(k)} = \frac{\gamma(0)}{2},
\end{equation}
up to $O(a^2)$ corrections, while $\alpha_m^{(k)} = O(a^{m-2}) \rightarrow 0$ for $m \geq 3$: the $a^m$ of the step beats the $1/(4a^2)$ of the matrix element, so only drift and diffusion survive the continuum limit, and only the $\omega \rightarrow 0$ behavior of the spectral function enters. Equation \eqref{eq:kramers_moyal} therefore converges to Eq.~\eqref{eq:fokker-plank}.

\emph{Derivation of Eq.~\eqref{eq:phig_diffusion}.} Define the marginal distribution of the global phase,
\begin{equation}
    p(\phi_g, t) \equiv \int d\vec{\phi}\;\, \delta\Big(\phi_g - \tfrac{1}{N^2}\textstyle\sum_k \phi_k\Big)\, P(\vec{\phi}, t),
\end{equation}
differentiate in time, insert Eq.~\eqref{eq:fokker-plank}, and integrate by parts in each $\phi_k$ (no boundary terms on the torus). Each derivative acting on the $\delta$-function obeys $\partial_{\phi_k}\delta = -\tfrac{1}{N^2}\,\partial_{\phi_g}\delta$, since the $\delta$-function depends on $\phi_k$ only through the average. The drift contribution then appears in the combination $\sum_k F_k = \sum_k \partial E/\partial\phi_k \equiv 0$, which vanishes identically at every configuration ($E$ depends only on phase differences), so no assumption about the shape of $P$ is needed. For the diffusion term, a second integration by parts gives $N^2$ identical terms, each carrying two factors of $1/N^2$:
\begin{equation}
    \frac{\partial p}{\partial t} = \frac{\gamma(0)}{4} \times N^2 \times \frac{1}{N^4}\, \frac{\partial^2 p}{\partial \phi_g^2},
\end{equation}
which is Eq.~\eqref{eq:phig_diffusion}. The reduction is exact: the flat direction closes on itself because the drift cancellation holds pointwise.

\emph{Derivation of Eq.~\eqref{eq:Tbf_fp}.} Let $\tau(x)$ be the mean first-exit time from the interval $(-\pi/2, \pi/2)$ for a walker starting at $\phi_g = x$. Over a small step $dt$ the walker moves by $\delta x$ with $\langle \delta x \rangle = 0$ and $\langle \delta x^2 \rangle = \tfrac{\gamma(0)}{2N^2}\,dt$ [from Eq.~\eqref{eq:phig_diffusion}], and, because the walk is Markovian, the remaining journey is a fresh first-passage problem from the new starting point:
\begin{equation}
    \tau(x) = dt + \big\langle \tau(x + \delta x) \big\rangle
    = dt + \tau(x) + \frac{\gamma(0)}{4N^2}\,\tau''(x)\, dt + \ldots
\end{equation}
Canceling $\tau(x)$ from both sides and dividing by $dt$ gives the backward-Kolmogorov equation quoted in the main text, $\tfrac{\gamma(0)}{4N^2}\,\tau''(x) = -1$, with absorbing boundary conditions $\tau(\pm\pi/2) = 0$ (a walker starting on the boundary has already exited). Integrating twice, using the $x \rightarrow -x$ symmetry to drop the linear term, yields the parabola of the main text, whose value at the center $x = 0$ is Eq.~\eqref{eq:Tbf_fp}.

\bibliography{refs}

\end{document}